\documentclass[11pt,twoside,letterpaper]{article} 
\usepackage{times,fancyhdr}
\usepackage{graphicx,amsmath,amssymb}
\usepackage{url}
\usepackage{wrapfig,subfigure}
\DeclareGraphicsExtensions{.pdf}
\makeatletter 
\def\cleardoublepage{\clearpage\if@twoside \ifodd\c@page\else%
    \hbox{}%
    \thispagestyle{empty}%
    \newpage%
    \if@twocolumn\hbox{}\newpage\fi\fi\fi} 
\makeatother 
\def\figurename{Figure}
\makeatletter
\renewcommand{\fnum@figure}[1]{\figurename~\thefigure.}
\makeatother
\def\tablename{Table}
\makeatletter
\renewcommand{\fnum@table}[1]{\tablename~\thetable.}
\makeatother
\def\be{\begin{equation}}
\def\ee{\end{equation}}
\def\ba{\begin{array}}
\def\ea{\end{array}}
\def\bea{\begin{eqnarray}}
\def\eea{\end{eqnarray}}
\def\bi{\begin{itemize}}
\def\ei{\end{itemize}}

\def\half{{\textstyle{1\over2}}}

\def\ve{\varepsilon}
\begin{document}
\title{
{\begin{flushleft}
\vskip 0.45in
{\normalsize\bfseries\textit{Chapter~1}}
\end{flushleft}
\vskip 0.45in
%
%
%
%
\bfseries\scshape The nuclear symmetry energy, the inner crust, and global neutron star modeling}}
\author{\bfseries\itshape W.G. Newton, M. Gearheart, J. Hooker and Bao-An Li \thanks{E-mail: william\_newton@tamu-commerce.edu}\\
Department of physics and astronomy, Texas A\&M University-Commerce,\\
Commerce, TX, USA}
\date{}
\maketitle
\thispagestyle{empty}
\setcounter{page}{1}
\thispagestyle{fancy}
\fancyhead{}
\fancyhead[L]{In: Neutron Star Crust \\ 
Editors: C.A. Bertulani and J. Piekarewicz, pp. {\thepage-\pageref{lastpage-01}}} 
\fancyhead[R]{ISBN 0000000000  \\
\copyright~2012 Nova Science Publishers, Inc.}
\fancyfoot{}
\renewcommand{\headrulewidth}{0pt}
\vspace{2in}

\pagestyle{fancy}
\fancyhead{}
\fancyhead[EC]{W.G. Newton}
\fancyhead[EL,OR]{\thepage}
\fancyhead[OC]{Symmetry energy, inner crust and global modeling}
\fancyfoot{}
\renewcommand\headrulewidth{0.5pt} 
%
\section{Introduction}

The outer crust of a neutron star, below densities of $\rho \sim 4 \times 10^{11}$ g cm$^{-3}$ consists of matter in a state not too far removed from that found in white dwarfs: a lattice of nuclei permeated by a relativistic, degenerate electron gas which gives the dominant contribution to the pressure of the matter \cite{BPS1971}. As pressure increases with depth, equilibrium with respect to weak interactions drives the nuclei to become more neutron rich. There comes a point when the intra-nuclear forces can no longer bind all the neutrons, and \emph{neutron drip} occurs. Above $\rho \sim 4 \times 10^{11}$ g cm$^{-3}$, a new regime is entered in which the nuclear lattice is bathed in a fluid of (`dripped') neutrons. These neutrons are delocalized much like conduction band electrons in metals. From this density inwards, the \textbf{equation of state} (\textbf{EOS}) is dominated by pressure arising from nucleon-nucleon interactions.

\begin{figure}[!tb]\label{fig:1}
\begin{center}
\includegraphics[width=14cm,height=8cm]{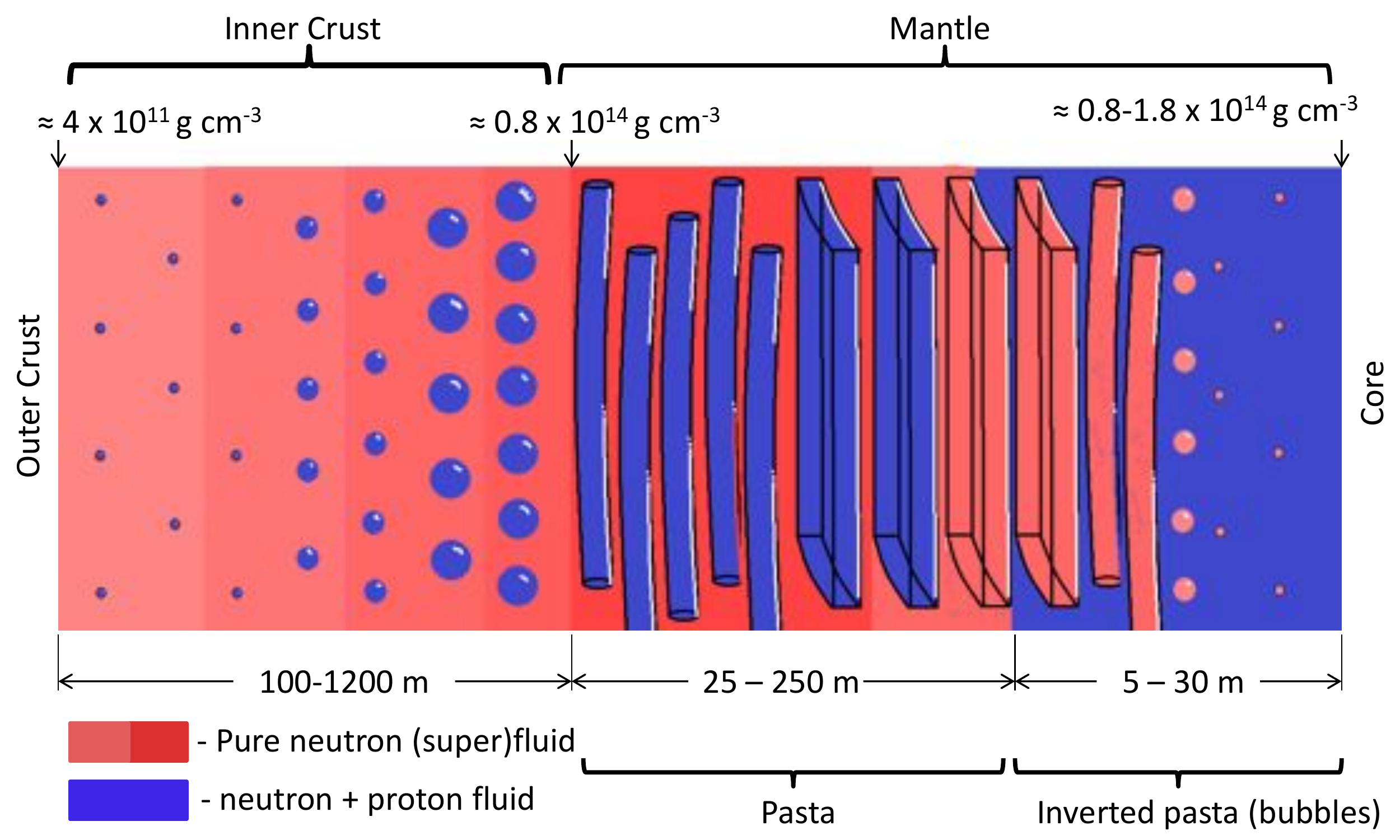}
\caption{Cartoon representation of the inner neutron star crust.}
\end{center}
\end{figure}

A cartoon representation of the crustal layers below the outer crust is shown in Fig.~1. Models predict two distinct layers. (1) The \textbf{inner crust} between densities $\rho \sim 4 \times 10^{11}$ g cm$^{-3}$ and $\rho \sim 10^{14}$ g cm$^{-3}$ is an elastic solid consisting of a lattice of heavy, neutron rich nuclei surrounded by fluid neutrons, with the nuclei increasing in size and mass with density while the inter-nuclear spacing and nuclear proton fraction decrease \cite{BBP1971}.  (The presence of a background electron gas will be taken as given from now on). The dripped neutrons are expected to become superfluid shortly after the neutron star is formed as it rapidly cools below temperatures of $10^8 - 10^9$ K \cite{Dean2003}. (2) The \textbf{mantle} \cite{Gusakov2004} between $\rho \sim 10^{14}$ g cm$^{-3}$ and the crust-core transition density consists of frustrated matter: the competition between the nuclear surface energy and the nuclear and lattice Coulomb energies over similar length scales drives the formation of exotic nuclear geometries termed nuclear `pasta'  \cite{Ravenhall1983, Oyamatsu1984} which proceed through a canonical sequence of phases: cylindrical (spaghetti) $\to$ slab (lasagna) $\to$ cylindrical bubble $\to$ spherical bubble. The latter three (`bubble') phases are distinguished by the delocalization of the \textbf{charged nuclear component} of the matter (containing the protons) in one or more dimensions, and corresponding localization of the \textbf{charge-neutral nuclear component} (fluid neutrons). Similar microscopic structures are observed in terrestrial soft condensed matter systems such as surfactants \cite{Jones2002,Watanabe2005}; by analogy, we can expect rich mechanical properties, intermediate between liquid and elastic solid, to emerge in the mantle \cite{Pethick1998}. Some crust models predict the absence of the mantle \cite{Douchin2001}; its presence depends sensitively on the nuclear microphysics of the crust. Fig.~1 gives a range of widths for the inner crust and mantle taken from the model presented in this chapter and encompassing a range of neutron star masses from 1-2 $M_{\odot}$ and a range of equations of state as discussed later.

To describe the states of matter in a neutron star, one needs a model for the nucleon-nucleon interactions as they are manifested in a many-nucleon context. A useful concept that bridges the gap between \emph{ab initio} nucleon-nucleon calculations, nuclear experimental observables, and neutron star matter is that of uniform \textbf{nuclear matter (NM)}. This is an idealized system, homogeneous and infinite in extent, of neutrons and protons interacting solely via the strong force. The energy per particle of such a system at a density $\rho$ and proton fraction $x$, $E(\rho, x)$, is referred to as the nuclear matter equation of state (\textbf{NM EOS}). In the regions of the neutron star core where protons and neutrons exist, the NM EOS can be combined with the electron energy, and under conditions of charge neutrality and beta-equilibrium gives an EOS for the core. In the inner crust, the NM EOS can be used to describe the dripped neutrons ($x=0$) and the bulk matter in the nuclear clusters. A consistent model for the EOS of crust and core necessarily uses a unique NM EOS, and one should expect parameters characteristic of a given NM EOS to correlate with both crust and core properties.

Nuclear matter with equal numbers of neutrons and protons ($x=0.5$) is referred to as \textbf{symmetric nuclear matter (SNM)}; nuclear matter with $x=0.0$ is naturally referred to as \textbf{pure neutron matter (PNM)}. Nuclei on Earth contain closely symmetric nuclear matter at densities close to nuclear saturation density $\rho_0 \approx 2.7 \times 10^{14}$ g cm$^{-3} \equiv 0.16$ fm$^{-3} = n_0$, where we use $n$ to refer to baryon number density. Thus experiment has constrained the properties of $E(\sim n_0,\sim0.5)$ to within relatively tight ranges, but the properties of PNM remain uncertain from an experimental standpoint. In the past decade, much experimental activity has been devoted to extending our knowledge of nuclear interactions to more neutron-rich systems and to higher and lower densities. Although we cannot produce pure neutron matter in the laboratory, we can produce matter with proton fractions as low as $x \approx 0.3$ in certain neutron rich isotopes and in the products of heavy ion collisions. This allows us to obtain information on how $E(\sim n_0, x)$ changes as $x$ decreases. 

By expanding $E(n, x)$ about $x = 0.5$ using the isospin asymmetry variable $\delta = 1-2x$, we can define a useful quantity called the \emph{symmetry energy} S(n),
\be\label{eq:eos1}
	E(n,\delta) = E_{\rm 0}(n) + S(n)\delta^2 + ...; \;\;\;\;\;\;\;\; S(n) = {1 \over 2}{\partial^2 E(n,\delta) \over \partial \delta^2}\bigg|_{\delta=0},
\ee
\noindent which encodes the change in the energy per particle of NM as one moves away from isospin symmetry. This allows extrapolation to the highly isospin asymmetric conditions in neutron stars. The simplest such extrapolation, referred to as the \textbf{parabolic approximation (PA)}, truncates the expansion to second order, giving
\be \label{PAapp}
E_{\rm PNM}(n) \equiv E(n, \delta=1) \approx E_{\rm 0}(n) + S(n)
\ee
for the PNM EOS. Expanding the symmetry energy about $\chi=0$ where $\chi = \frac{n-n_{\rm 0}}{3n_{\rm 0}}$ we obtain
\be\label{eq:eos3}
	S(n) = J + L \chi + \half K_{\rm sym} \chi^{2} + ..., 
\ee
\noindent where $J$, $L$ and $K_{\rm sym}$ are the symmetry energy, its slope and its curvature at saturation density. 

Since neutron star matter contains a low fraction of protons, many inner crust and global stellar properties are sensitive to the symmetry energy parameters $J$,$L$, etc. To give a simple example, the pressure of PNM at saturation density is given by $P_{\rm PNM}(n_0)$=$n_0L/3$. The pressure in the inner crust and outer core is dominated by neutron pressure so a strong correlation exists between the pressure in neutron stars near saturation density and $L$. Neutron star EOSs which have higher pressures are often referred to as `stiff'; lower pressure EOSs are referred to as `soft'. Thus, around 1 to 2$n_0$, `stiff' EOSs are associated with high values of $L$ and `soft' EOSs with low values of $L$.

That large uncertainties exist in characteristic NM parameters such as $J$ and $L$ is one reason why many model predictions of potential neutron star observables span such wide ranges. On the other hand, observations of neutron stars offer the opportunity to obtain constraints on NM parameters and hence on the underlying models of the nucleon-nucleon interactions \cite{Lattimer2001}.

The following is a non-exhaustive list of (potentially) observable neutron star phenomena whose precise properties depend on the properties of the inner crust and mantle and of the star as a whole; some of these will be described in detail elsewhere in this book.

\vspace{\baselineskip}

$\bullet$ \hspace{5pt} \textbf{Pulsar glitches.} Young pulsars spin down under the action of magnetic torque, their rotational energy powering the radiation beam. Many are observed to undergo occasional, sudden, spin-ups called glitches \cite{Espinoza2011}. Proposed mechanisms include crust-cracking as the star attempts to adjust its shape to become more spherical \cite{Baym1969} and angular momentum transfer from one internal component to another such as some part of the crust superfluid neutrons to the rigid part of the crust \cite{Anderson1975, Link1999}, or a combination of both \cite{Ruderman1998}. In such models, the size, frequency and post-glitch relaxation of the spin period depend on, among other microscopic properties, the crust and core sizes, moments of inertia and composition.

$\bullet$ \hspace{5pt} \textbf{Free precession.} Certain pulsars exhibit long timescale periodic variation in their timing residuals, with periods of order years, suggestive of free precession of the star \cite{Stairs2000,Shabanova2001}. Free precession is can arise from mechanically or magnetically supported crustal deformation \cite{Cutler2002,Wasserman2003,Cutler2003}, and the period depends also on details of crust-core coupling, notably through the properties of the crust and core superfluid \cite{Jones2001,Link2003,Link2006,Glampedakis2009}.

$\bullet$ \hspace{5pt} \textbf{QPOs from SGR giant flares.} Quasi-periodic oscillations (QPOs) in the tails of light curves of giant flares from soft gamma-ray repeaters (SGRs) have been observed  \cite{Israel2005, Watts2006, Strohmayer2005, Strohmayer2006}, and their frequencies lie in the range of possible torsional vibrations of the crust. The crust thickness, composition (through, e.g., the shear modulus) and the stellar size all affect the frequencies of such modes \cite{SteinerWatts2009,Samuelsson2007,Andersson2009,Gearheart2011,Sotani2011}.

$\bullet$ \hspace{5pt} \textbf{Neutron star cooling.} The crust thermalization timescale depends on the crust thickness as well as the thermal conductivity and specific heats arising from the heat transport mechanisms operating in the crust \cite{Lattimer1994, Gnedin2001}. One intriguing possibility is the operation of the direct Urca process in the bubble phases of the mantle, where the delocalization of the protons may allow it \cite{Gusakov2004}. The thickness of this layer plays an important role in determining how effective a cooling mechanism this might be.

$\bullet$ \hspace{5pt} \textbf{Gravitational waves (GWs) from neutron stars.} A rich array of stellar oscillation modes are possible, some of which might generate GWs detectable on Earth \cite{Andersson2011,Abbott2010}, or lead to other observational signatures such as limiting neutron star spin-up \cite{Bildsten1998,Andersson1999}. Stability of modes can depend sensitively on the physics at the crust core interface and crust thickness, \cite{Bildsten00,Andersson00,Lindblom00,Rieutord01,Peralta06,Glampedakis06,Wen11}. GWs can also be generated by a quadrupole deformation in the stellar shape supported, among other possibilities, by the elastic crust \cite{Ushomirsky2000, Haskell2006}. Whether the crust is strong enough to support a large enough deformation to produce detectable gravitational waves depends on the shear modulus throughout the crust, and thus its composition (especially in the mantle where the mechanical properties are particularly uncertain) and the crust thickness and stellar size \cite{Gearheart2011}.
\vspace{\baselineskip}

In this chapter we will review the dependences of the composition, thickness of the crust and the mantle, and certain global neutron star properties on the symmetry energy parameters $J$ and $L$. The interplay of such relationships in modeling neutron star observables and the potential for obtaining astrophysical constraints on $J$ and $L$ will be illustrated by constructing consistent crust and core models based on a model of uniform nuclear matter whose symmetry energy parameters can be smoothly varied. We shall use the \textbf{compressible liquid drop model (CLDM)} of crustal matter for its expediency; an outline of more sophisticated models will be given in the discussion at the end of the chapter.

\section{Symmetry energy and its correlations with observables}

In this section we review experimental and theoretical constraints on $J$ and $L$ and some of the correlations that emerge between $J$ and $L$ and neutron star observables. Throughout this chapter we will illustrate the correlations using a particular model for nuclear matter: the phenomenological modified Skyrme-like (MSL) model \cite{MSL01,LieWenChen2010}. The energy as a function of density $n$ and isospin asymmetry $\delta$ is written down in a form that closely resembles the uniform nuclear matter Skyrme EOS under the Hartree-Fock approach. The advantage of the MSL function is that its free parameters can be more easily related to the properties of nuclear matter at saturation density and allows for a smooth variation of $J$ and $L$ independently while keeping the SNM EOS constant. The MSL EOS is written
\be \label{eq:emsl}
E^{\rm MSL}(n,\delta) = \frac{\eta}{n}\left(\frac{\hbar^2}{2m_n^*}n_n^{5/3}+\frac{\hbar^2}{2m_n^*}n_p^{5/3}\right) + \frac{\alpha}{2}\frac{n}{n_0} + \frac{\beta}{\sigma+1}\frac{n^{\sigma}}{n_0^{\sigma}} + E_{sym}^{loc}(n)\delta^2
\ee
\be\label{eq:emsloc}
	E_{sym}^{loc}(n) = (1-y)E_{sym}^{loc}(n_0)\frac{n}{n_0} + yE_{sym}^{loc}(n_0)\left(\frac{n}{n_0}\right)^{\gamma_{\rm sym}},
\ee
\noindent where 
\be\label{effmass1}
{\hbar^2 \over 2m_n^*} = {\hbar^2 \over 2m} + n(C_{\rm eff} + D_{\rm eff} \delta); \;\;\;\; {\hbar^2 \over 2m_p^*} = {\hbar^2 \over 2m} + n(C_{\rm eff} - D_{\rm eff} \delta),
\ee
\noindent $\eta = (3/5) (3\pi^2)^{2/3}$, $n_0$ is the saturation density and $\alpha$, $\beta$, $\sigma$, $C_{\rm eff}$, $D_{\rm eff}$, $\gamma_{\rm sym}$, $E_{\rm sym}^{\rm loc}(n_0)$ and $y$ are free parameters. $C_{\rm eff}$ and $D_{\rm eff}$ are fixed by setting the effective masses at saturation to be $m_{\rm p,0}^* = 0.8m$ and $m_{\rm n,0}^*=0.7m$, where $m$ is the average nucleon mass in free space. We set  $\gamma_{\rm sym}$ = 4/3; $\alpha$, $\beta$, $\sigma$ are set by the incompressibility $K_0$, density $n_0$ and energy per particle $E_0$ of SNM at saturation, while  $E_{\rm sym}^{\rm loc}$ and $y$ control the absolute value of the symmetry energy and its slope respectively. We will keep constant $n_0 = 0.16$ fm$^{-3}$, $E(n_0, x=0.5) = -16$ MeV and $K_0$ = 240 MeV. The symmetry energy in the MSL model is not restricted by the parabolic approximation (Eq. (2)), but includes contributions of orders higher than $\delta^2$ from only the kinetic and effective mass parts;  the potential part of the symmetry energy, $E_{sym}^{loc}(n)$, is quadratic in isospin asymmetry.

\subsection{Experimental constraints on the symmetry energy}

\begin{figure}[!t]\label{fig:2}
\begin{center}
\includegraphics[width=7.0cm,height=6cm,trim=0 0 0 16]{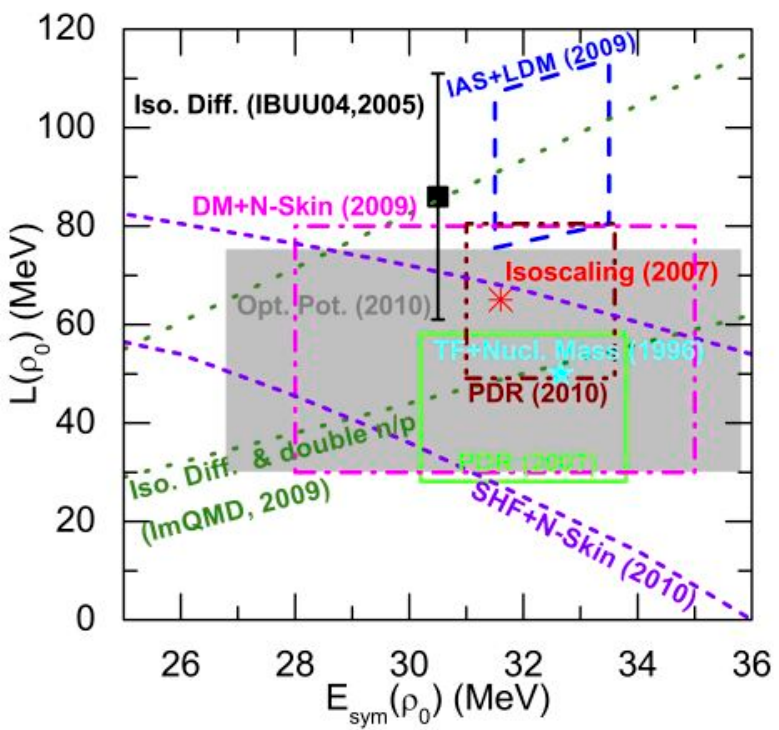}\includegraphics[width=7.0cm,height=6.5cm,trim=0 0 0 0]{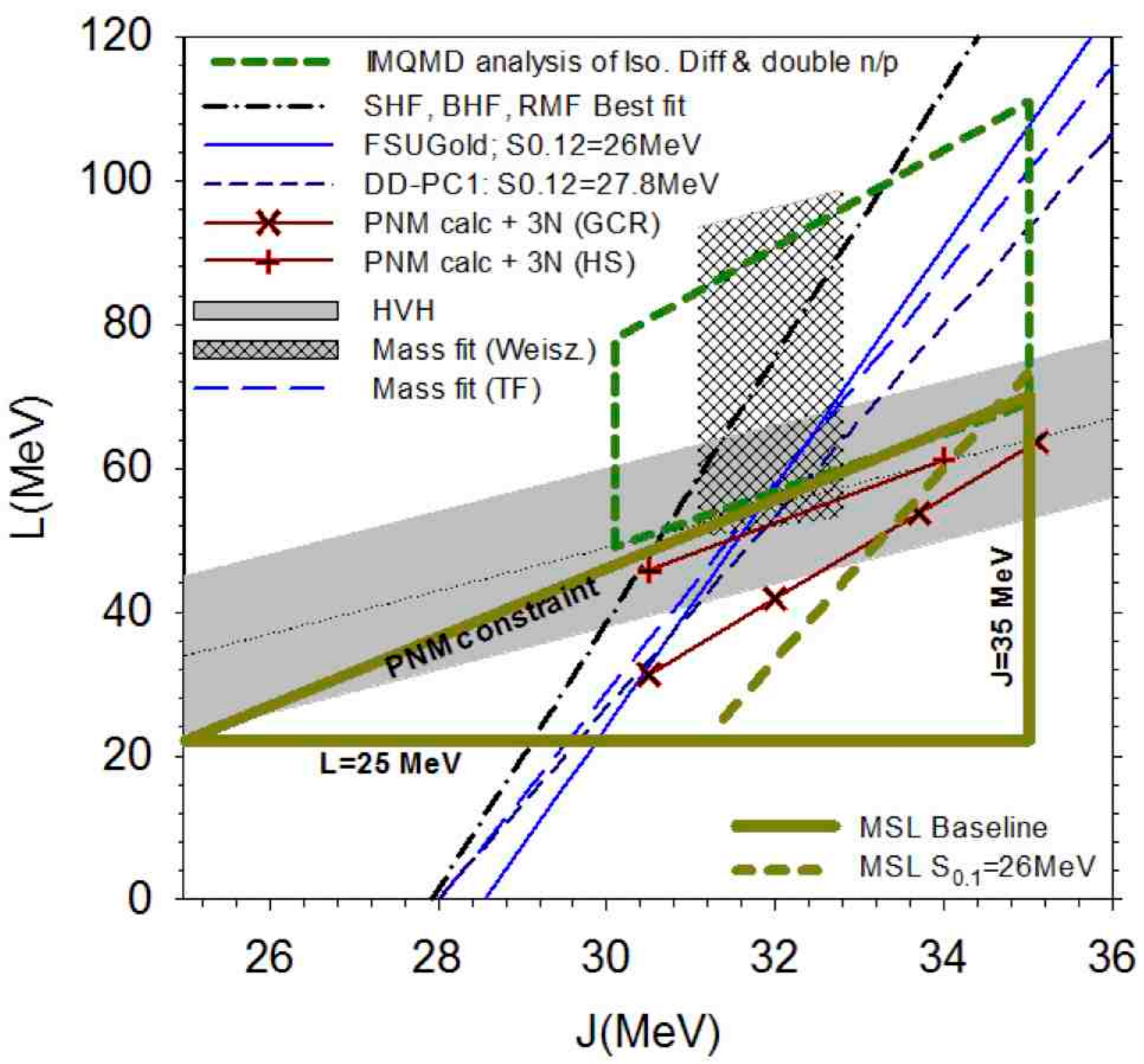}
\caption{\textbf{Left panel}: Symmetry energy slope at saturation density $L$ versus its magnitude $E_{\rm sym}(\rho_0) \equiv J$ extracted from isospin diffusion and neutron/proton (n/p) ratios of pre-equilibrium nucleon emissions within the Improved Molecular Dynamics (ImQMD-2009) model \cite{Tsang2004,Tsang2009,Famiano2006}, isospin diffusion within the Isospin-Dependent Boltzmann-Uehling-Ulenbeck model (IBUU04- 2005) \cite{Chen2005,Li2005}, isoscaling \cite{Shetty2007}, energy shift of isobaric analogue states within the liquid drop model (IAS+LDM-2009) \cite{Danielewicz2009}, neutron(n)-skins of several heavy nuclei using the droplet model (DM) or the Skyrme-Hartree-Fock (SHF) approach \cite{LieWenChen2010,Centelles2009,Warda2009}, pygmy dipole resonances (PDR) in 68Ni,132Sn and 208Pb, \cite{Klimkiewicz2007,Carbone2010}, and nucleon global optical potentials (GOP)~\cite{ChangXu2010}. Taken from ref.~\cite{ChangXu2010}. \textbf{Right panel}: Correlations between $J$ and $L$ arising from the MSL EOS constrained by $J$=35 MeV, $L$=25 MeV, low density PNM calculations (together outlining our `baseline' region) and $S$(0.1fm$^{-3}$), fits to nuclear masses using the Weisz\"acker~\cite{Liu2010} and Thomas-Fermi (TF)~\cite{Oyamatsu2010} models, the Hugenholtz-van-Hove (HVH) theorem applied to global optical potentials~\cite{ChangXu2010} from which the overall bounds on the left plot were extracted, PNM calculations taking into account uncertainties in 3-nucleon (3N) interactions using chiral perturbation theory~\cite{Hebeler2010} and quantum Monte-Carlo (QMC) simulations~\cite{Gezerlis2008,Gandolfi2011}, the relativistic mean field (RMF) FSUGold model constrained by $S$(0.12fm$^{-3}$)=26MeV~\cite{FSUGold,Fattoyev2010}, the RMF DD-PC1 model constrained by $S$(0.12fm$^{-3}$)=27.6MeV~\cite{Moustakidis2010}, a best fit to a large number of SHF, Bruekner-Hartree-Fock (BHF) and RMF models \cite{Ducoin2011}, and the model correlations from the ImQMD-2009 analysis of isospin diffusion and n/p ratios from which the overall bounds on the left plot were extracted \cite{Tsang2009}. Adapted from \cite{Newton2011}.}
\end{center}
\end{figure}

The current status of our uncertainty in the symmetry energy based on a wide range of experimental results is summarized in the left panel of Fig.~2; the caption summarizes the details. We point out again that the parameters $J$ and $L$ describe an idealized system, and the extracted values are somewhat model dependent. However, by using many independent observables we can check the consistency of our results and arrive at a more robust set of constraints. We will take as a conservative range from experiment $25\textless L \textless 115$ MeV, noting however that a convergence of experimental results to the lower half of this range is apparent in Fig.~2. For $J$ we take 25$\textless J \textless$35 MeV, which includes the ranges extracted from nuclear mass model fits \cite{Liu2010,Myers1966,Moller1995,Pomorski2003}.

\subsection{Theoretical constraints on the symmetry energy}

Symmetry energy constraints follow from theoretical calculations of the PNM EOS. At low densities ($\lesssim$ 0.02fm$^{-3}$) where only two-nucleon (2N) interactions need be considered, quantum Monte-Carlo (QMC), Green's function Monte-Carlo, effective field theory (EFT), chiral EFT and variational chain summation techniques \cite{Hebeler2010,Gezerlis2008,Gandolfi2011,Akmal1998,Carlson2003,Schwenk2005} have produced robust constraints on the PNM EOS which may be taken as a constraint for our MSL model. In Fig.~3 we show the PNM constraint from Schwenk and Pethick (SP) \cite{Schwenk2005} as the red box at low densities; results since have converged on the lower bound of that box. Calculations at higher sub-saturation densities which require 3-nucleon (3N) interactions to be included start to diverge in their predictions; two recent calculations with estimated theoretical error bars for the 3N forces have been performed \cite{Hebeler2010, Gandolfi2011} (hereafter labelled HS, GCR respectively), and are indicated by the shaded bands in Fig.~3.

\begin{figure}[!t]\label{fig:3}
\begin{center}
\includegraphics[width=7.3cm,height=6cm]{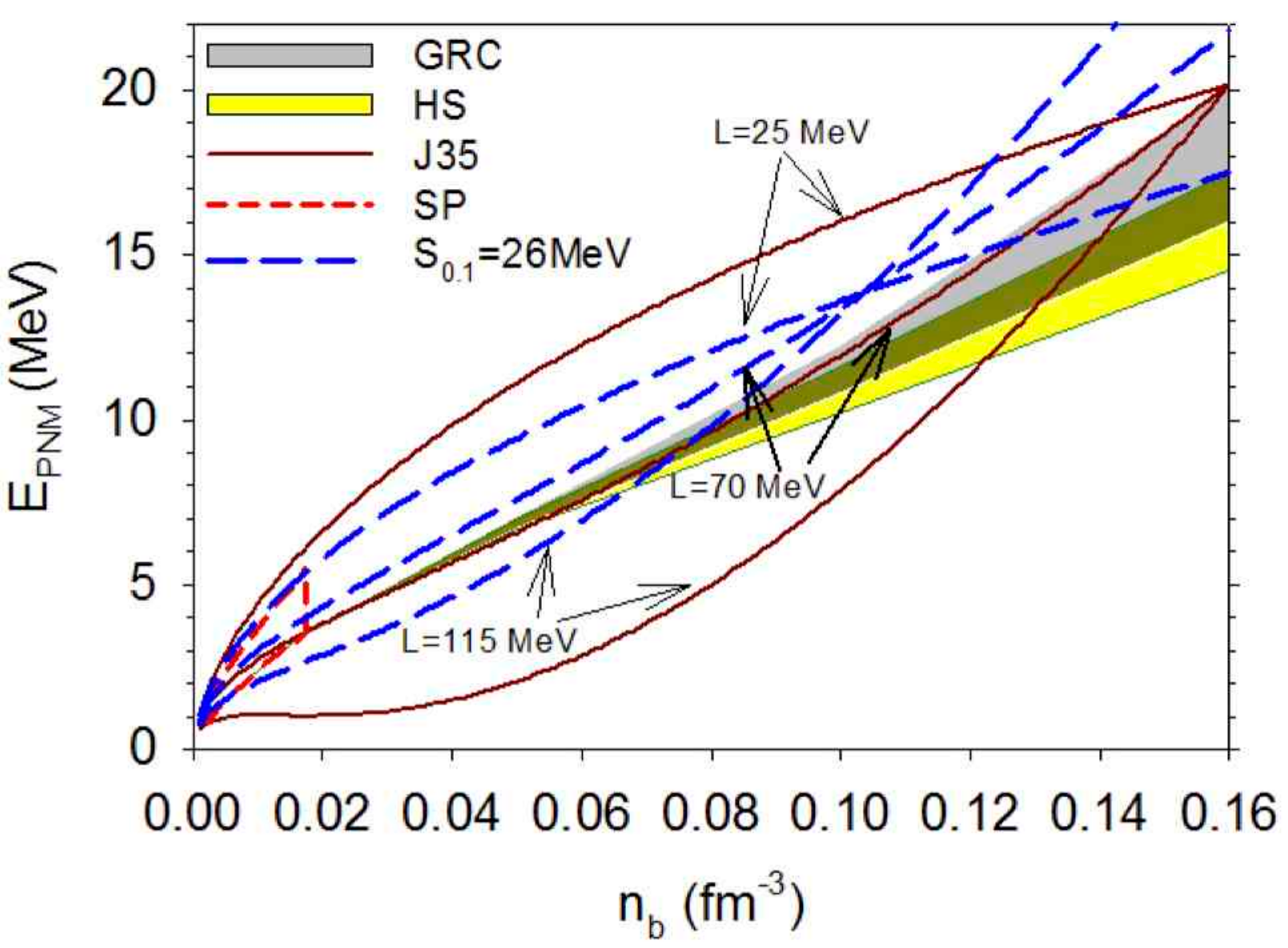}\includegraphics[width=6.7cm,height=6cm]{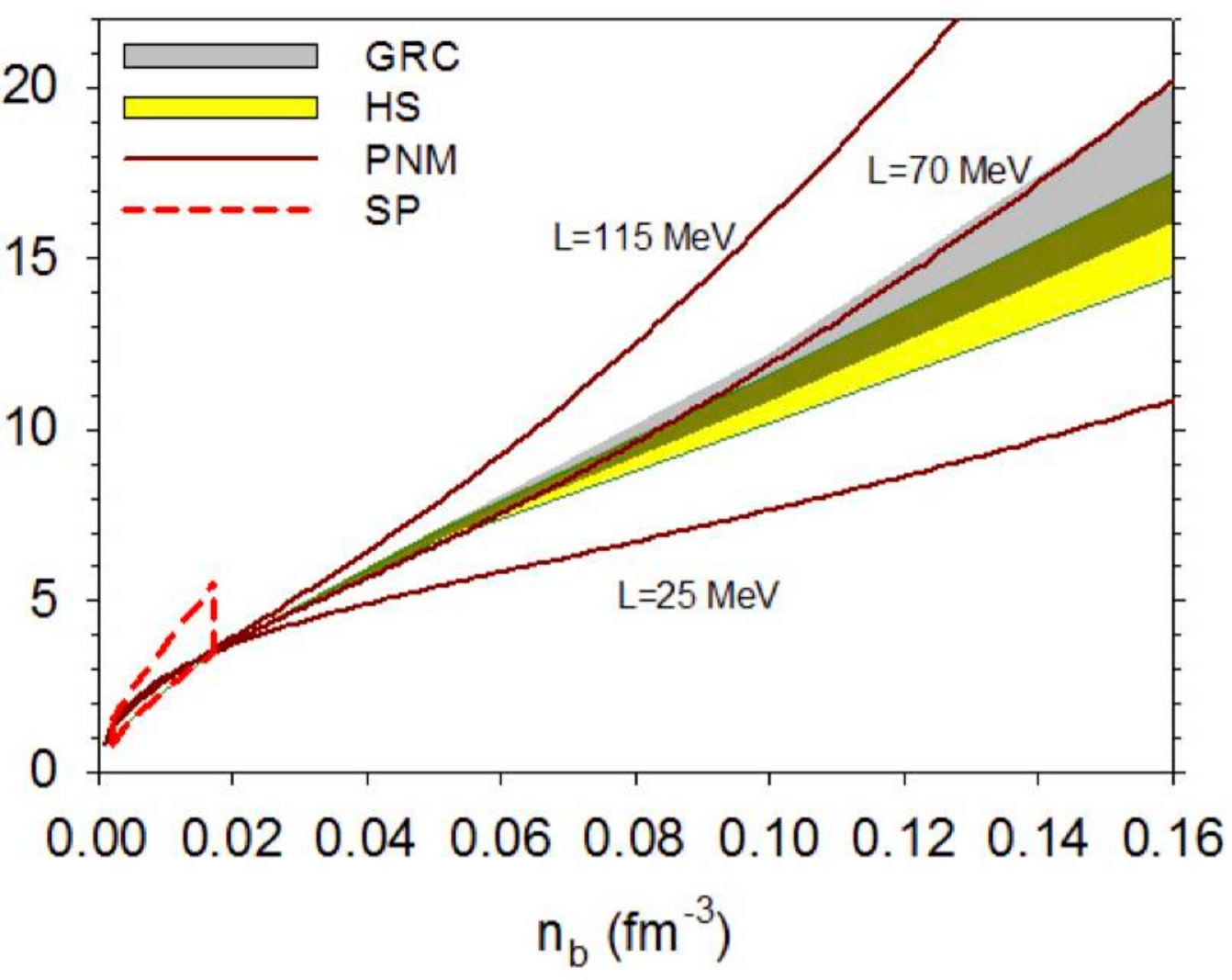}
\caption{Pure neutron matter (PNM) energy per particle versus baryon number density from effective field theory (EFT) at low densities (SP) \cite{Schwenk2005}, chiral EFT with 3-nucleon (3N) interations (HS) \cite{Hebeler2010} and quantum Monte-Carlo (QMC) with 3N interactions (GCR) \cite{Gezerlis2008,Gandolfi2011}. The MSL EOS is plotted with symmetry energy slopes of $L=25,70,115$ MeV and with $J=35$ MeV (`J35'), $S(0.1 {\rm fm}^{-3})$ = 26 MeV (`S$_{0.1}$') (\textbf{left panel}) and constrained to fit the low-density PNM EOS of SP, GCR and HS (`PNM') (\textbf{right panel}). Adapted from \cite{Newton2011}.}
\end{center}
\end{figure}

\subsubsection{Correlations between $J$ and $L$}

In the left panel of Fig.~3, we plot the PNM EOS from our MSL model for $J=35$ MeV and $L=25, 70, 115$ MeV. A fixed $J$ corresponds closely to a fixed value of the PNM EOS  $E(n_0, x=0.0)$ at saturation density for a fixed SNM EOS (see Eq.~(2)). Only the $L = 70$ MeV EOS agrees with the predictions of theory at low densities. However, one can adjust $J$ for a given value of $L$ to obtain agreement with the low density PNM EOS; the results for the MSL EOS are displayed in the right panel of Fig.~3. Doing so naturally introduces correlation between $J$ and $L$; in the right panel of Fig.~2 we display the correlation obtained in this way for the MSL model. It is fit by $J = 20.53 + 0.207L$. For reference, the correlations obtained directly from the PNM calculations of HS and GCR, using the PA (Eq. (2)) with $E_0=-16$ MeV to obtain $J$ from $E_{\rm PNM} (n_0)$, are depicted in Fig.~3; although offset slightly from the MSL results, their slopes are similar. A similar correlation is obtained from the Hugenholtz-Van-Hove (HVH) theorem which predicts a relation between $J$ and $L$ whose uncertainty can be related to global nucleon optical potentials \cite{ChangXu2010}

One experimental probe of the symmetry energy is the measurement of neutron skins of nuclei. This probes the symmetry energy at densities around $n=0.1$fm$^{-3}$; thus many models fix the symmetry energy at this density. In the right panel of Fig.~3 we show the MSL PNM EOSs constrained by $S(0.1$fm$^{-3}$) = 26 MeV; varying $L$ then produces a steeper correlation with $J$, also shown in the right panel of Fig.~2; $J = 29.0 + 0.1L$. It is worth noting that increasing the density at which one fixes the symmetry energy in a given model, increases the slope in the $J$-$L$ plane.

Similar correlations are obtained from two relativistic mean field models \cite{Fattoyev2010, Moustakidis2010} and from a best fit to a wide selection of model predictions of $J$ and $L$ \cite{Ducoin2011}, also shown in the left panel of Fig.~2. Finally we also show correlations that emerge from nuclear mass fits \cite{Liu2010,Oyamatsu2010} and analysis of data from heavy ion collisions \cite{Tsang2009}.

In what follows we shall use sequences of MSL EOSs generated by varying $L$ with a variety of constraints on $J$: the sequence generated keeping $J$ fixed will be labelled, e.g., `\textbf{J35}'; the sequence generated by fixing the low density PNM EOS will be labelled the `\textbf{PNM}' sequence; and the sequence generated by fixing $S(0.1$fm$^{-3}$) = 26 MeV will be labelled the `$\bf{S_{0.1}}$' sequence. The model correlations in the right panel of Fig.~2 overlap in the region $25\textless L\textless70$ MeV, in line with the most recent experimental results. By combining the MSL `PNM' constraint with the requirement that $25\textless J \textless35$ MeV and $L\textgreater$25MeV we obtain a region in the $J$-$L$ plane which we shall refer to as our `\textbf{baseline}' region.

\subsection{Correlations with neutron star properties}

Some useful correlations of symmetry energy parameters with basic neutron star properties have been established, which we review here; more details can be found in the following references: \cite{Lattimer2001,Fattoyev2010,Ducoin2011,Steiner2005,Ravenhall2004}

\vspace{\baselineskip}

\noindent $\bullet$ \hspace{5pt} \textbf{ The pressure of neutron star matter} in beta-equilibrium at $n_0$ including the electron contribution can be approximated \cite{Lattimer2001,Ravenhall2004}
\be \label{PNS}
P_{\rm NS} (n_0) \approx {n_0 \over 3} L + 0.048 n_0 \bigg( {J \over 30} \bigg)^3 \bigg(J - {4 \over 3} L\bigg),
\ee
\noindent where the second term provides a correction of only 2-3\% for $L=25$ MeV, rising to 10-20\% for  $L=115$ MeV, with $J$ over the range 25 - 35 MeV. At densities slightly above or below this, extra terms are introduced, but the leading order will remain the one proportional to $L$ alone.

\noindent $\bullet$ \hspace{5pt} \textbf{The radius} of a neutron star is found to correlate with the pressure at a fiducial density inside the star \cite{Lattimer2001} of the form
\be\label{RP}
R \approx C(M,n) P(n)^{0.25},
\ee
\noindent where the pressure in MeV fm$^{-3}$ is taken at a density between $n_0$ and $2n_0$. $C(M,n)$ is a constant for a given stellar mass $M$ and fiducial density $n$. Given Eq. (\ref{PNS}) to leading order and taking $n = n_0$, this suggests \cite{Newton2009b}
\be\label{RL}
R \approx 0.5 C(M, n_0) L^{0.25},
\ee
\noindent using $(n_0/3)^{0.25} \approx 0.5$. For a $1.4 M_{\odot}$ star, $C(M,n_0) \approx 9.3$. \textbf{The compactness parameter $\beta = GM/Rc^2$} thus goes as
\be\label{RL}
\beta \approx {2 G M \over C(M, n_0) c^2} L^{-0.25}.
\ee
	
\noindent $\bullet$ \hspace{5pt} \textbf{Total moment of inertia of the star} $I$ scales with $MR^2$ \cite{Lattimer2001,Ravenhall2004}, so
\be\label{I}
I \sim M  C(M, n_0) L^{0.5},
\ee

\noindent $\bullet$ \hspace{5pt} \textbf{The crust-core transition density $n_{\rm cc}$} has been established to correlate inversely with the slope of the symmetry energy \cite{Ducoin2011,Oyamatsu2007, Steiner2008, Xu2009}; if all other nuclear matter parameters are held constant, the correlation is generally found to be weakly parabolic:
\be \label{ncc1}
n_{\rm cc} = a_{\rm n} - b_{\rm n} L + c_{\rm n} L^2,
\ee
\noindent where $a_{\rm n}\sim0.1 {\rm fm}^{-3},  b_{\rm n}\sim10^{-3}{\rm fm}^{-3} {\rm MeV}^{-1}, c_{\rm n}\sim4\cdot10^{-6} {\rm fm}^{-3} {\rm MeV}^{-2}$ depend on other model parameters for their precise values \cite{FuturePaper}. For $L\sim$10 MeV, the quadratic term is negligible, and the relation is closely linear; it becomes important for $L\sim$100 MeV.

\noindent $\bullet$ \hspace{5pt} \textbf{The inner crust-mantle transition density $n_{\rm p}$}, that is, the density at which pasta appears, is roughly constant with respect to the symmetry energy parameters \cite{Oyamatsu2007}, with the exact value depending on the inner crust model. Typically,
$n_{\rm p} \sim 0.04 - 0.06\; {\rm fm}^{-3}.$

\noindent $\bullet$ \hspace{5pt} \textbf{The crust-core transition pressure $P_{\rm cc}$} was initially though to correlate with $L$ in a similar way to $n_{\rm cc}$, as expected from Eq.~(\ref{PNS}); however, a more thorough survey of nuclear matter models reveal no significant correlation with $L$ or $J$ \cite{Fattoyev2010,Ducoin2011}. This can be understood by the fact that although the pressure at  a given density correlates positively with $L$, the transition density correlates negatively, and the convolution of the two correlations \emph{for a given nuclear matter model} could be positive or negative depending on the other nuclear matter parameters. Over a wide range of NM models, therefore, there will be no obvious correlation. However, \cite{Ducoin2011} did find a robust correlation relating $P_{\rm cc}$ to symmetry energy parameters at densities characteristic of $n_{\rm cc}$:
\be \label{Pcc}
P_{\rm cc} = a_{\rm p} + b_{\rm p} [L(0.1 {\rm fm}^{-3}) - 0.343 K_{\rm sym} (0.1 {\rm fm}^{-3})],
\ee
\noindent where $a_{\rm p}\sim0.5 {\rm MeV fm}^{-3},b_{\rm p}\sim0.01 {\rm fm}^{-3}$ depend on other model parameters for their precise values \cite{Ducoin2011,FuturePaper}. Note also that $P_{\rm cc}$ correlates linearly with the baryon chemical potential at the crust-core transition $\mu_{\rm cc}$

\noindent $\bullet$ \hspace{5pt} \textbf{The relative crust thickness, mass and moment of inertia} go as \cite{Lattimer2001,Fattoyev2010,Ravenhall2004}
\be\label{DR}
{\Delta R \over R}  \hspace{-3pt} \sim  \hspace{-3pt} {R \over M} \mu_{\rm cc} \hspace{-3pt} \sim  \hspace{-3pt} {C \over M} L^{0.25} \mu_{\rm cc}; \;\; {\Delta M \over M}  \hspace{-3pt} \sim  \hspace{-3pt} {R^2 \over M^2} P_{\rm cc}  \hspace{-3pt} \sim  \hspace{-3pt} {C^2 \over M^2} L^{0.5} P_{\rm cc};  \;\; {\Delta I \over I}  \hspace{-3pt} \sim  \hspace{-3pt} {R^4 \over M^2} P_{\rm cc}  \hspace{-3pt} \sim  \hspace{-3pt} {C^4 \over M^2} L P_{\rm cc}. \nonumber
\ee

\noindent Higher order corrections to global crust and star properties depending on radius will be ordered in ascending powers of $\beta$; higher order corrections to global crust properties will be additionally ordered in ascending powers of $P_{\rm cc}/\epsilon_{\rm cc}$ or equivalently $\mu_{\rm cc}/m_{\rm n}$ where $\epsilon_{\rm cc}$ is the mass-energy density at the crust-core transition and $m_{\rm n}$ is the neutron rest mass.

\section{The Compressible Liquid Drop Model}

In order to examine the symmetry energy dependence of inner crustal properties, we will use the simple, but physically insightful, compressible liquid drop model (CLDM) for the energy density of crustal matter. The CLDM formalism which we highlight below was originally formulated by BBP \cite{BBP1971} and updated to incorporate pasta shapes \cite{Iida1997,Watanabe2000}. Its computational expediency makes it widely used as a model for the crust EOS (e.g. \cite{Douchin2001,LorenzPethick1993}) and suitable for a calculation of  many crust compositions and EOSs over a range of nuclear matter parameters. Its simplicity is also its disadvantage, neglecting as it does important microphysics; these problems will be discussed later in the chapter. For further details of the model see \cite{BBP1971,Newton2011,Watanabe2000}.

A unit cell of crust matter is approximated by an equally volumed cell with the geometry of the nuclear cluster under scrutiny; this is the Wigner-Seitz (WS) approximation. The dimensionality of the shapes is specified by a parameter $d=3, 2, 1$ for spherical, cylindrical and planar geometries respectively. The total energy density of the matter an be written (neglecting rest masses)
\be\label{eq:ecell}
	\ve_{\rm cell}(r_{\rm c},x,n,n_{\rm n}) = v \big[ n E(n,x) + \ve_{\rm ex} + \ve_{\rm th} \big] + u(\ve_{\rm surf} + \ve_{\rm (C+L)}) + (1-v) n_{\rm n} E(n_{\rm n}, 0) + \ve_{\rm e}(n_{\rm e}),
\ee
\noindent where $r_{\rm c}$ is the radius of the WS cell, $u = (r_{\rm N} / r_{\rm c})^d$ is the volume fraction occupied by the nuclei or the bubbles of radii (or half-width in the case of slabs) $r_{\rm N}$, $x$ and $n$ are the proton fraction and baryon density of the charged nuclear component and $n_{\rm n}$ the baryon density of the neutron fluid. $n_{\rm e}$ is the number density of electrons. Charge neutrality demands $n_{\rm e}= vnx$ where $v$ is the volume fraction of the charged nuclear component, defined as
\be\label{eq:v}
	v = \left\{
  \begin{array}{l l}
	\displaystyle u & \quad {\rm nuclei}\\
	\displaystyle 1-u& \quad{\rm bubbles},\\
  \end{array} \right.
\ee
\noindent and the global baryon number density is related to the local baryon densities through $n_{\rm b} = vn + (1-v)n_{\rm n}$. The contributions to the energy density of the cell break down as follows:

$\bullet$ \hspace{5pt} The electron kinetic energy density, that of an ultra-relativistic free Fermi gas, is $\ve_{\rm e} = (3/4)\hbar c k_{\rm e} n_{\rm e}$ with $k_{\rm e} = (3\pi^2 n_{\rm e})^{1/3}$.

$\bullet$ \hspace{5pt} The electrostatic contributions include the nuclear and lattice Coulomb energy densities, collectively written within the WS approximation as
\be\label{eq:ecl}
    \ve_{\rm (C+L)} = 2\pi (exnr_{N})^{2}f_{d}(u); \;\;\;  f_{d}(u) = \frac{1}{d+2} \left[\frac{2}{d-2}\left(1-{du^{1-2/d}\over 2} \right) + u\right].
\ee
\noindent The corrections to the nuclear Coulomb energy density from the finite surface thickness and the proton Coulomb exchange energy are given by
\be\label{eq:ethick}
	\ve_{th}(k,x) = -\frac{4}{9}\pi e^2w^2x^2k^3n; \;\;\;\;\; \ve_{ex}(k,x) = -\frac{3}{4\pi}2^{1/3}e^2x^{4/3}kn,
\ee
\noindent where $k = (1.5\pi^2 n)^{1/3}$ and $w$ is a distance representing the surface thickness, taken to be $w \approx 0.75$ fm. Electron screening can be included but is neglected in what follows.

$\bullet$ \hspace{5pt} The energy per particle of the neutron fluid and of neutrons and protons in the bulk of the nuclear clusters, $E(n,x)$ and $E(n_{\rm n},0)$ are obtained using a model for uniform nuclear matter. We will use the MSL model outlined in the previous section.

$\bullet$ \hspace{5pt} The surface energy density $\epsilon_{\rm surf}$ can be written in terms of the surface and curvature tensions $\sigma_{\rm s}$, $\sigma_{\rm c}$ as
\be
    \ve_{surf} = d \sigma_{\rm s} / r_{\rm N} + d(d-1) \sigma_{\rm c}/r_{\rm N}^2,
\ee
Thermodynamic equilibrium constrains the surface and curvature tensions to be a function of only one free parameter, conveniently taken to be $x$ \cite{Ravenhall1983.2, Lattimer1985}. One possible functional form is  \cite{Ravenhall1983.2, Lattimer1985,Lorenz1991} 
\be \label{eqn:surf}
    \sigma_{\rm s}(x)= \sigma_0 { 2^{p+1} + b \over {1 \over x^p} + b + {1 \over (1-x)^p} }; \;\;\;\;\;  \sigma_{\rm c}(x) = \sigma_{\rm s} {\sigma_{\rm 0,c} \over \sigma_{0}} {\alpha(\beta - x) + \gamma x^4}.
\ee
The parameters $\sigma_0, \sigma_{\rm 0,c}, b, p, \alpha, \beta, \gamma$ are typically adjusted to fit nuclear masses \cite{Moller1995,Steiner2008} or microscopic calculations of the energy of the interface between two phases of semi-infinite nuclear matter \cite{Lorenz1991,Douchin2000} using the same nuclear model responsible for the bulk nuclear matter terms in the CLDM. Finally, a description of a neutron skin can be consistently included in the CLDM \cite{Lattimer1985,Lorenz1991,Douchin2000,Danielewicz2003}.

The composition of the cell is obtained by minimizing the energy density of the unit cell with respect to the free parameters (e.g. $x,n,n_{\rm n}$ and $r_{\rm c}$). This produces four equations to be solved which correspond physically to mechanical, chemical and beta equilibrium of the cell plus the nuclear virial relation $\epsilon_{\rm (C+L)} = 2\epsilon_{\rm surf} + \epsilon_{\rm curv}$ which expresses the scaling between Coulomb, surface and curvature energy densities under equilibrium with respect to variation of the volume fraction of the charged nuclear component \cite{BBP1971}.

\section{Crust-core and inner crust-mantle transition}

\begin{figure}[!h]\label{fig:4}
\begin{center}
\includegraphics[width=7cm,height=5cm]{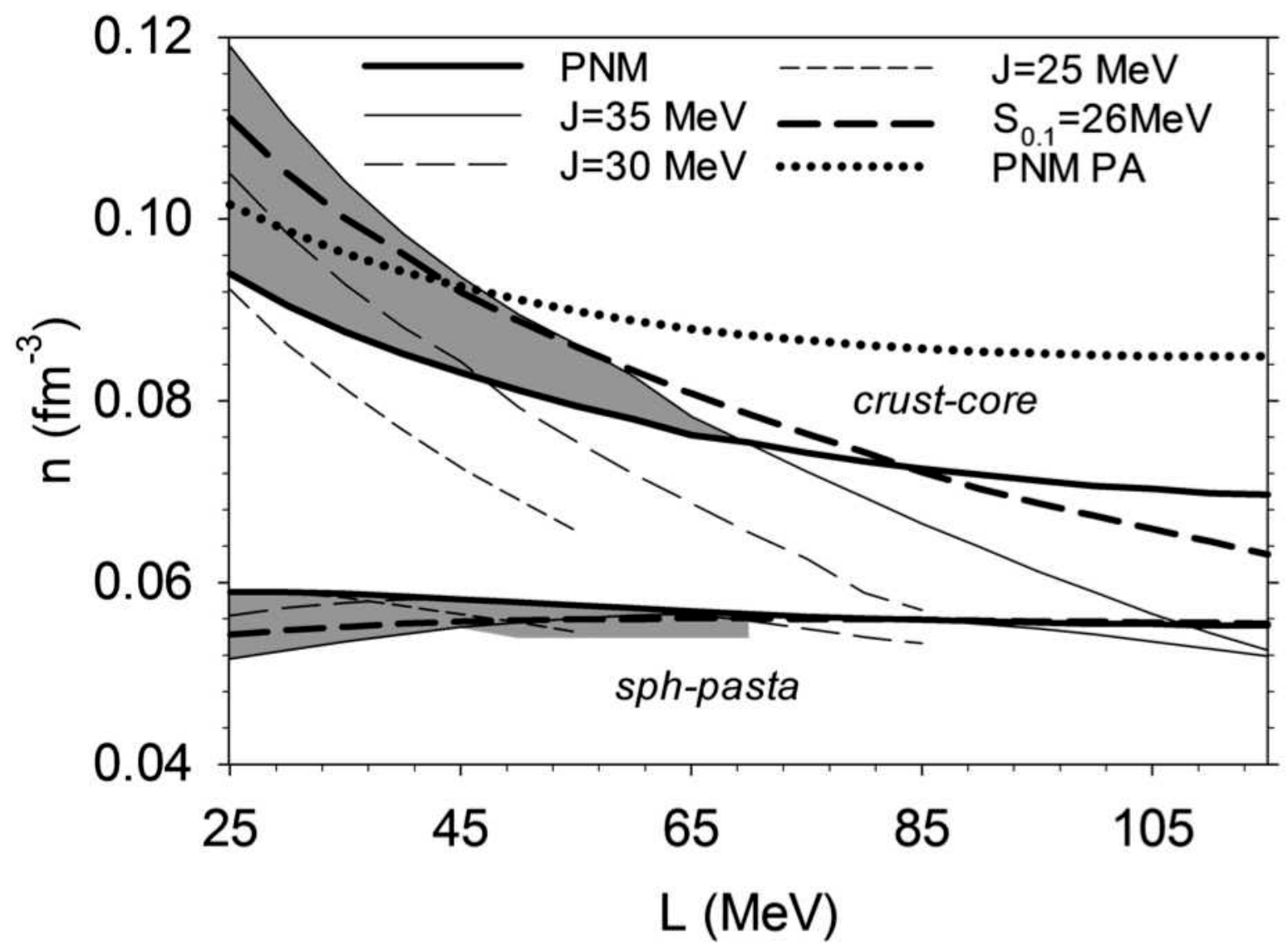}\includegraphics[width=7cm,height=5cm]{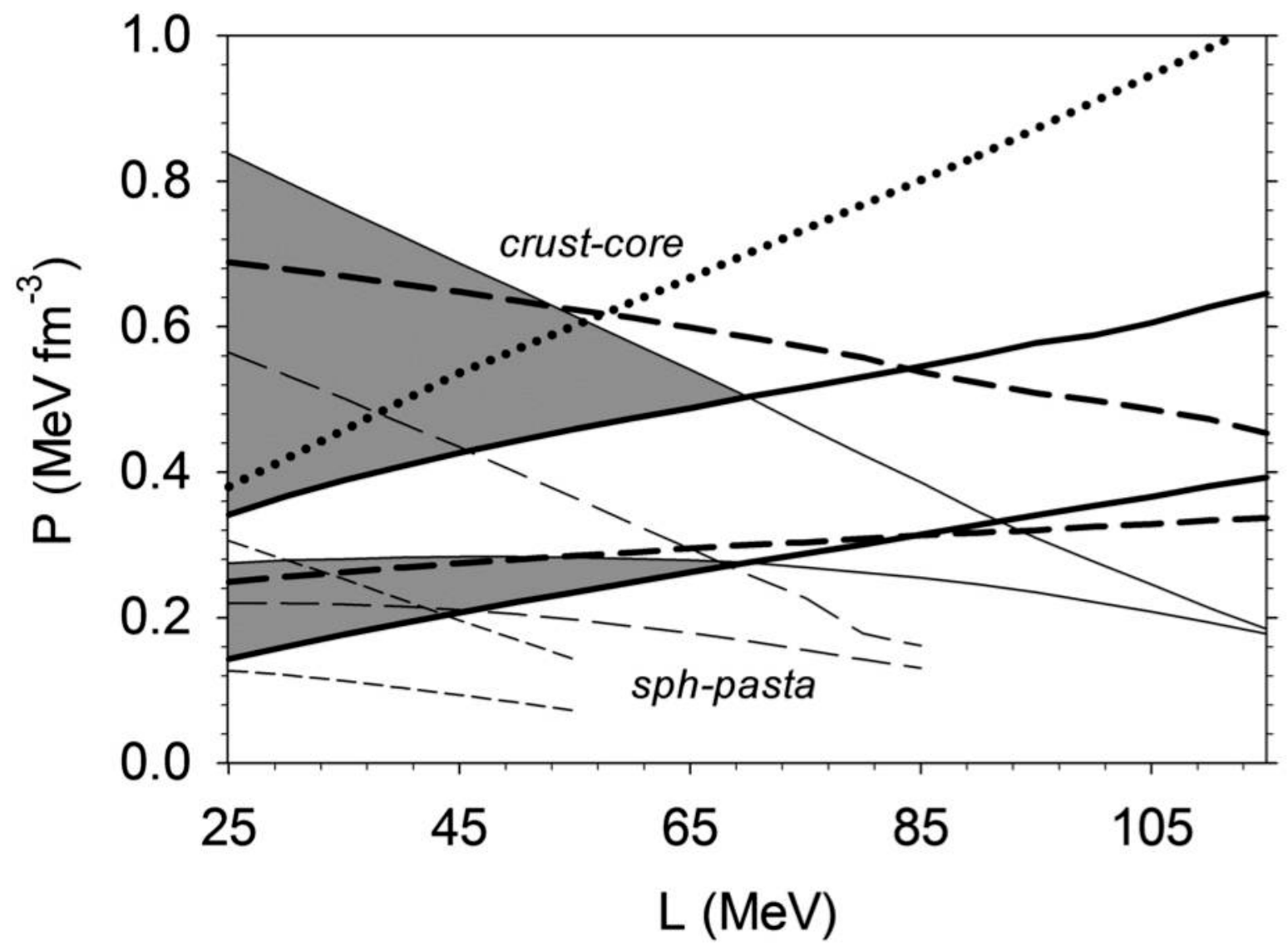}
\caption{Crust-core and spherical nuclei-pasta transition densities (left) and pressures (right) versus $L$ for the 3 constant-$J$ sequences `J35', `J30' and J25' and the sequences `PNM' and `S$_{0.1}$'. The effect of using the parabolic approximation (PA) to calculate crust-core transition properties for the `PNM' sequence are also shown. The shaded region indicates the baseline results. Adapted from \cite{Newton2011}.}
\end{center}
\end{figure}

We will now employ the CLDM formalism outlined in the previous section to calculate sequences of crust compositions and crust and core EOSs using the MSL nuclear matter model with $L$ varied between 25 and 115 MeV. The crust EOS sequences will be labelled according the corresponding nuclear matter EOS sequences outlined in Section~2.2.1: we shall use `J25', `J30', `J35',`PNM' and S$_{0.1}$ sequences. Results corresponding to the baseline region (Fig.~2) in $J$-$L$ space will appear shaded in the plots that follow.

The surface and curvature energy parameters in Eqs.~\ref{eqn:surf} should be obtained either from nuclear mass fits or microscopic calculations for each different set of $J,L$ values used. We adopt a different approach: we use the established correlation between the the surface symmetry energy (encoding the change in the surface energy in nuclei as they move away from isospin symmetry) and $J$ (see, e.g., \cite{Steiner2005}) to fix $b$ for a given $J$. Additionally we take $\sigma_0 = 1.1$MeV fm$^{-2}$, $p=3$ and $\sigma_{\rm 0,c}  = 0.6$MeV fm$^{-1}$ as typical values from microscopic calculations (see discussion in \cite{Newton2011}). The neutron skin is neglected.

In Fig.~4 we show the variation of the crust-core and inner crust-mantle ($\equiv$ spherical nuclei-pasta) transition densities and pressures for the five sequences of MSL EOSs. For the crust-core transition we also show results using the full parabolic approximation for the symmetry energy. We can summarize these results as follows:

$\bullet$ \hspace{5pt} \emph{The J25, J30, and J35 sequences}. For constant $J$, the crust-core transition density $n_{\rm cc}$ correlates negatively with $L$ as expected from previous studies. $n_{\rm cc}$ varies by $\approx 0.06$fm$^{-3}$ in the range $25\textless L \textless 115$ MeV. For constant $J$, the crust-core transition pressure $P_{\rm cc}$ correlates negatively with $L$, varying by $\approx 0.6$ MeV fm$^{-3}$ over the same range. $n_{\rm cc}$correlates positively with $J$, varying by $\approx 0.04$fm$^{-3}$ over $25\textless J \textless 35$ MeV for constant $L$. A higher symmetry energy favors a larger proton fraction in the nuclear clusters, lowering their bulk binding energy more than their surface energy is raised, thereby making the clustered matter energetically favorable to higher densities. $P_{\rm cc}$ also correlates positively with $J$ for fixed $L$, varying by $\approx 0.5$ MeV fm$^{-3}$ over  $25\textless J \textless 35$ MeV.

$\bullet$ \hspace{5pt} \emph{The PNM and S$_{0.1}$ sequences}. When a linear, positive correlation of $L$ with $J$, $L(J)$, is imposed, the relationship of $n_{\rm cc}$ and $P_{\rm cc}$ with $L(J)$ convolves their relationships with $L$ and $J$ indepedently. Fig~4 demonstrates the evolution of the $n_{\rm cc}$-$L$ and $P_{\rm cc}$-$L$ trends as the slope of $L(J)$ steepens from the `J35' sequence through `S$_{0.1}$' to `PNM';  the negative slope of the correlations get less pronounced, and, in the case of $P_{\rm cc}$, becomes positive for the steeper $L(J)$ slope of the `PNM' sequence. 

$\bullet$ \hspace{5pt} \emph{The parabolic approximation}. The results for the PA diverge to higher transition densities as $L$ increases compared to the full EOS. The MSL model is already parabolic in the potential part of the symmetry energy; our results thus demonstrate the importance of using the full kinetic part of the symmetry energy (as has been noted before \cite{Ducoin2011,Xu2009}).

$\bullet$ \hspace{5pt} Our baseline region gives a range of crust-core transition densities $n_{\rm cc}$=$0.08 - 0.12$ fm$^{-3}$ and pressures $P_{\rm cc} \approx 0.35 - 0.85$ MeV fm$^{-3}$.

$\bullet$ \hspace{5pt} Relative to the crust-core transition density, the density of transition to the pasta phases $n_{\rm p}$ shows little variation with $L$ and $J$, tracing out a thin band $n_{\rm p} \approx 0.05-0.06$ fm$^{-3}$; thus, the thickness of the pasta layers will correlate with $L$ and $J$ in a similar way to the crust-core transition density. The spherical-pasta transition pressure $P_{\rm p}$ increases with $L$ for the `PNM' sequence, whereas the variation with $L$ at constant $J$ is weaker. The baseline region gives $P_{\rm p} \approx 0.15-0.3$ MeV fm$^{-3}$.

\begin{figure}[!t]\label{fig:5}
\begin{center}
\includegraphics[width=7.0cm,height=5cm]{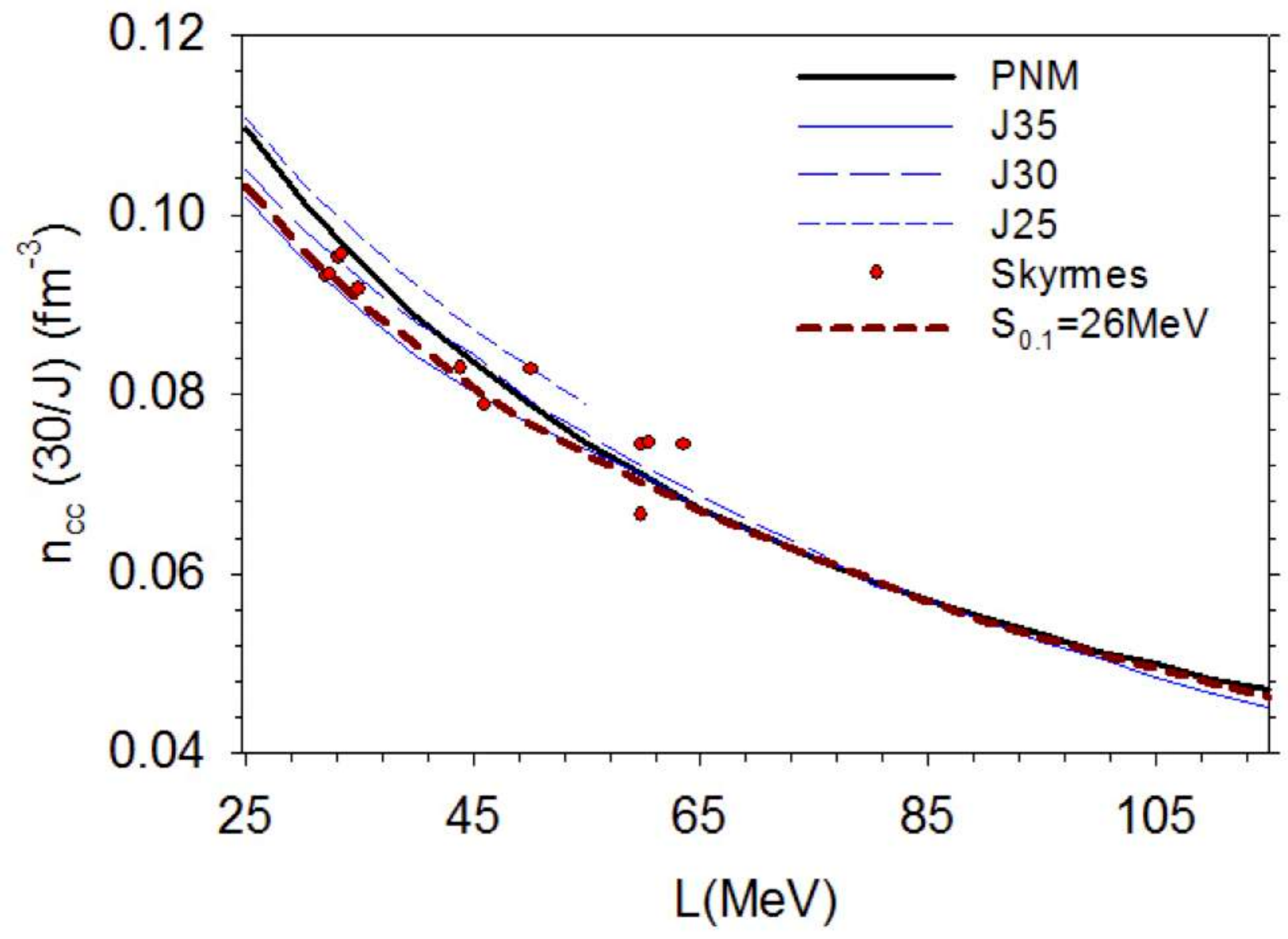}\includegraphics[width=7.0cm,height=5cm]{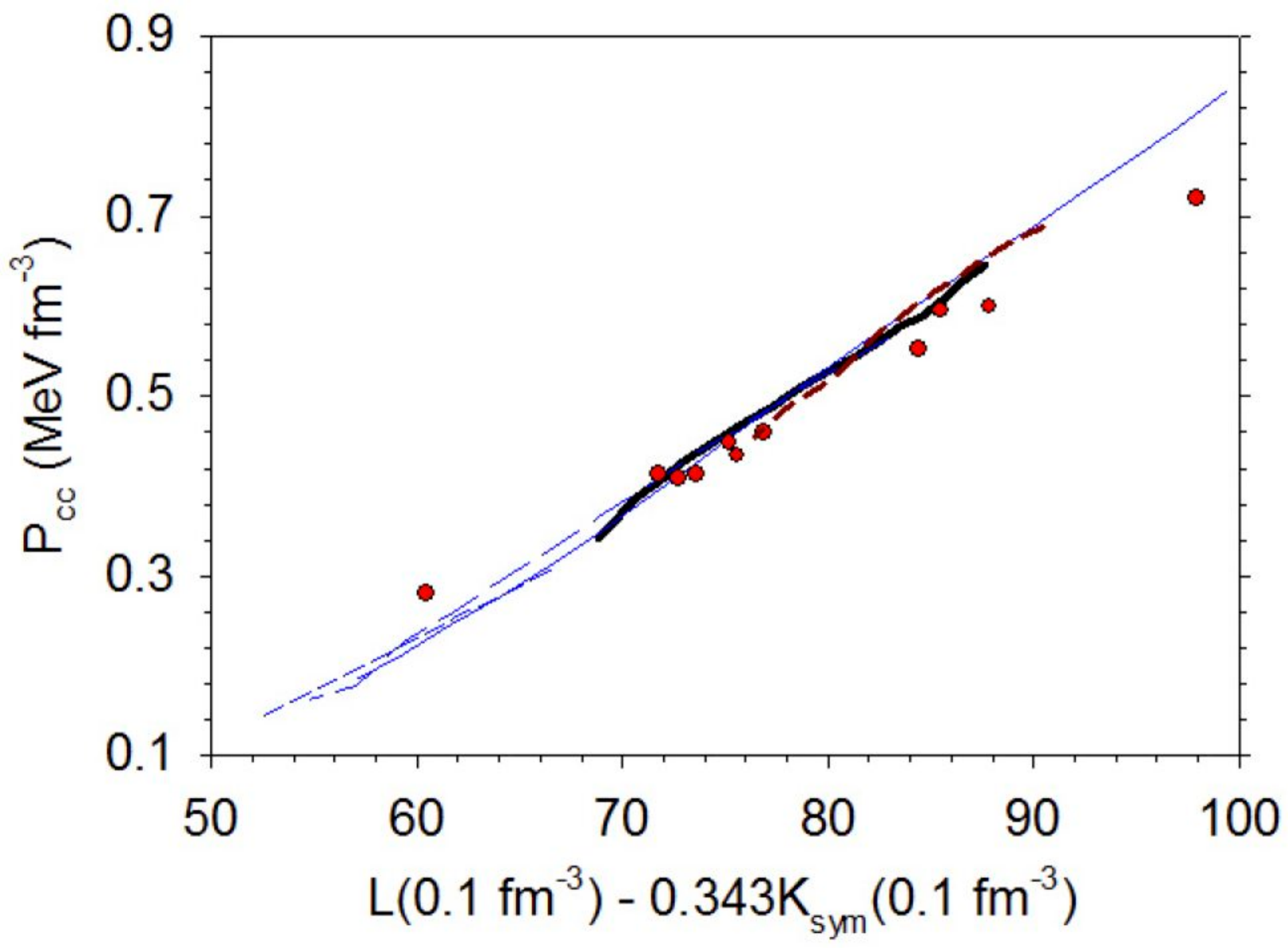}
\caption{Crust-core transition densities scaled by 30 MeV/$J$ (left) versus $L$ and transition pressures versus $L(n=0.1$ fm$^{-3}$) - 0.343 $K_{\rm sym}(n=0.1 $fm$^{-3}$) for the EOS sequences shown in Fig.~4. Adapted from \cite{Newton2011}.}
\end{center}
\end{figure}

These results demonstrate that the crust core transition density and pressure are sensitive not only to the slope of the symmetry energy $L$ at saturation but also to its magnitude $J$ to a similar extent over the experimentally constrained ranges; thus, quite different results can be obtained from calculations using different NM models which adopt, implicitly or explicitly, different correlations between $J$ and $L$.

The transition densities for the sequence of constant $J$ EoSs (Fig.~4) suggests a roughly inverse scaling of the densities with $J$. With this in mind, we plot in the left panel of Fig.~5 the transition densities multiplied by $30$MeV$/J$. The results form a relatively tight band correlated with $L$; included on the plot, as well as our MSL EOSs, are the results using a selection of Skyrme EOSs (see \cite{Newton2011} for details). Similarly, using the suggested scaling given in Eq.~(\ref{Pcc}) \cite{Ducoin2011}, the right panel of Fig.~5 shows a tight correlation, reinforcing this result and indicating that more experimental data on the symmetry energy parameters and including the curvature $K_{\rm sym}$ would improve our estimate of the transition pressure, as well as hinting at the connection between the $J$-$L$ correlations and the higher order symmetry parameters $K_{\rm sym}$, etc.

\section{Crust composition}

\begin{figure}[!t]\label{fig:6}
\begin{center}
\includegraphics[width=7cm,height=4.65cm]{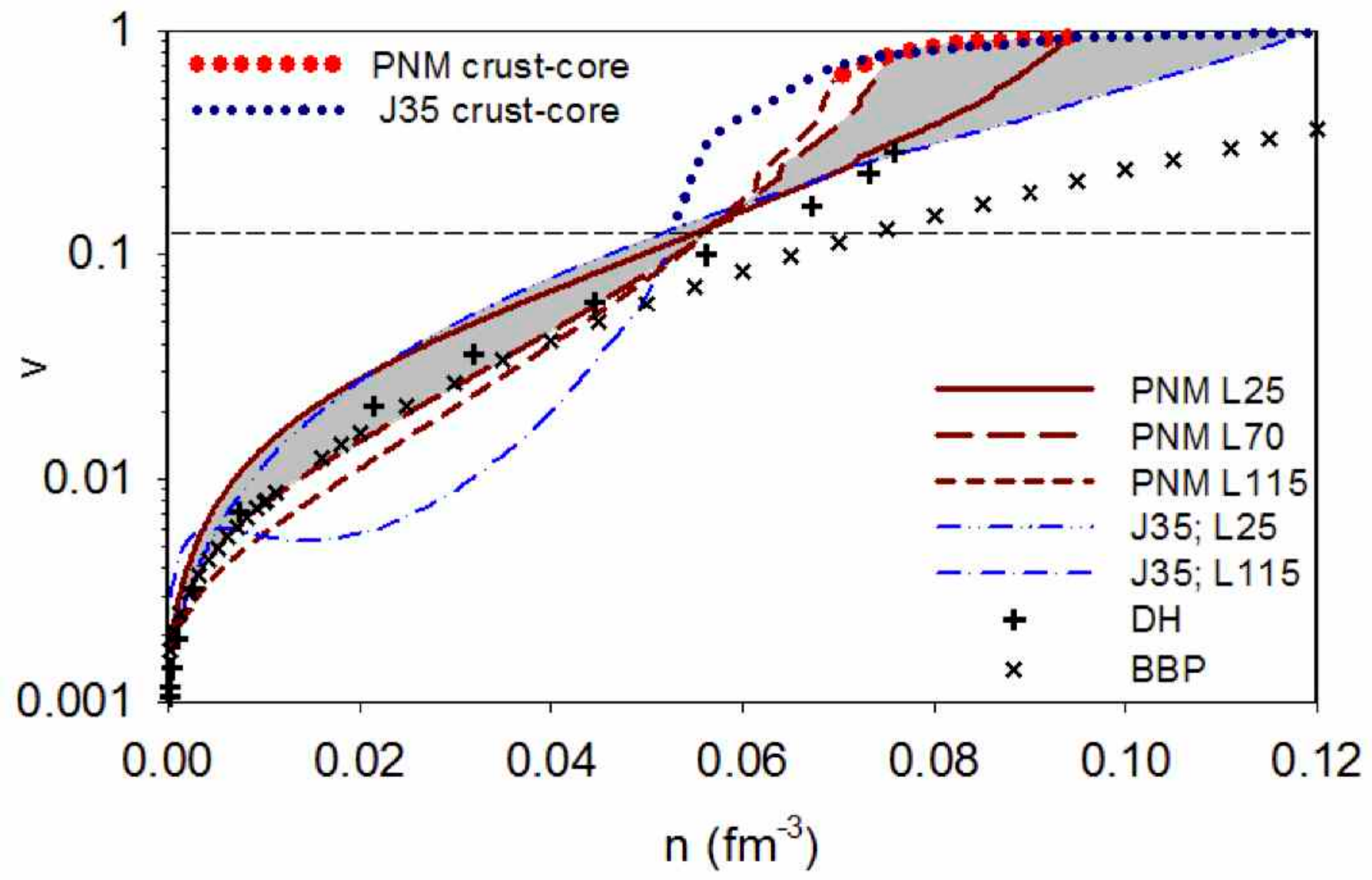}\includegraphics[width=7cm,height=4.65cm]{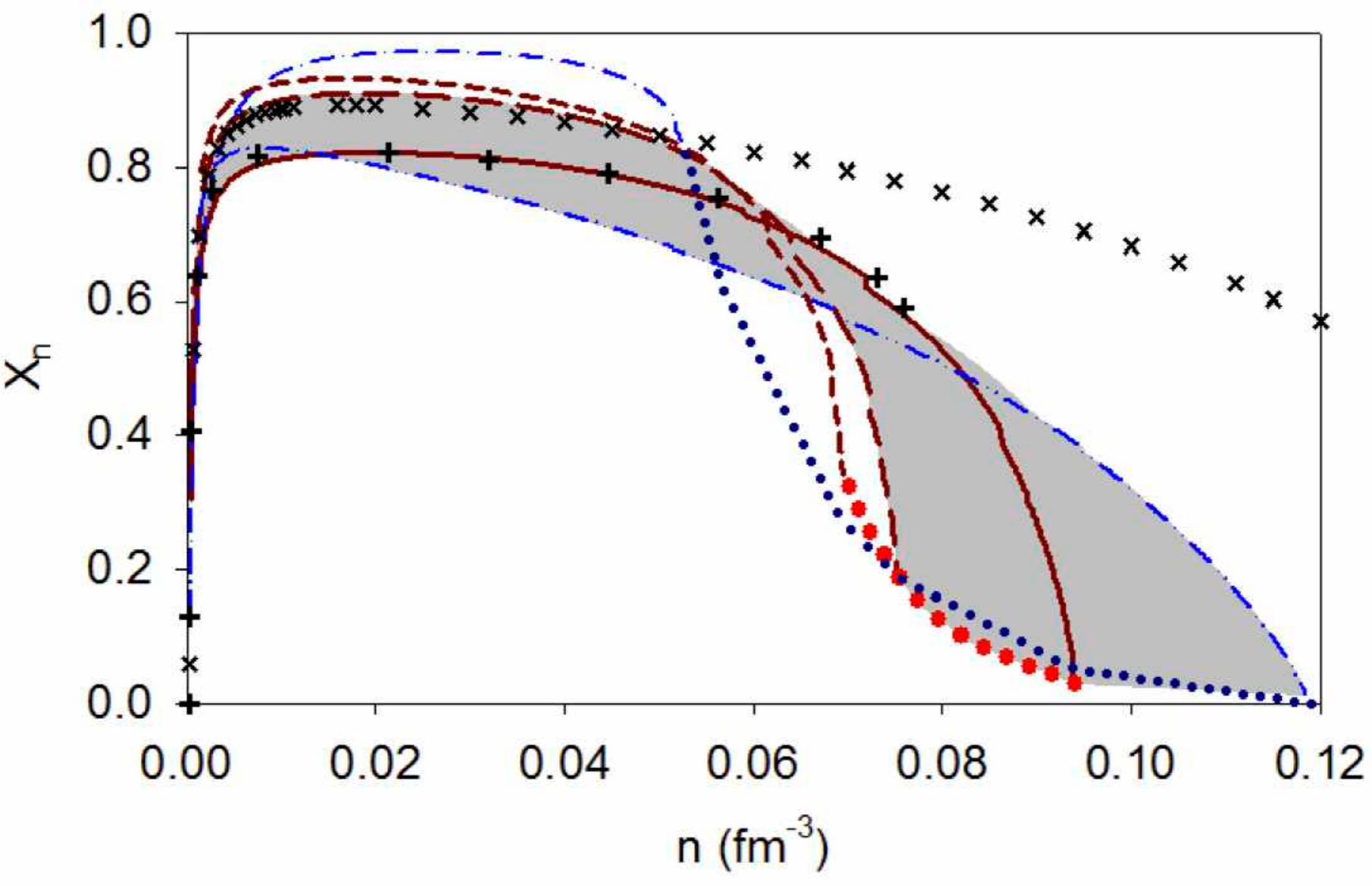}
\includegraphics[width=7cm,height=4.65cm]{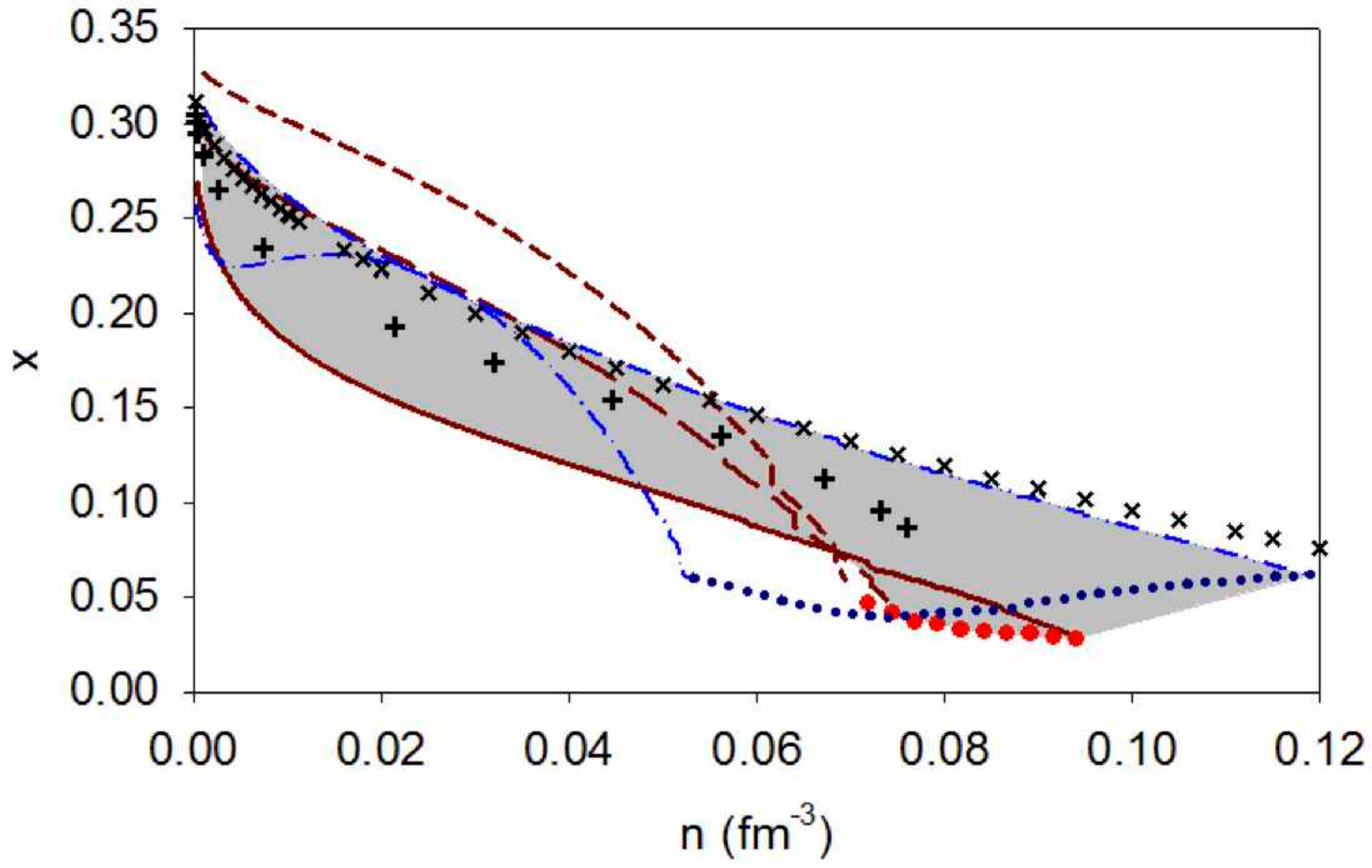}\includegraphics[width=7cm,height=4.65cm]{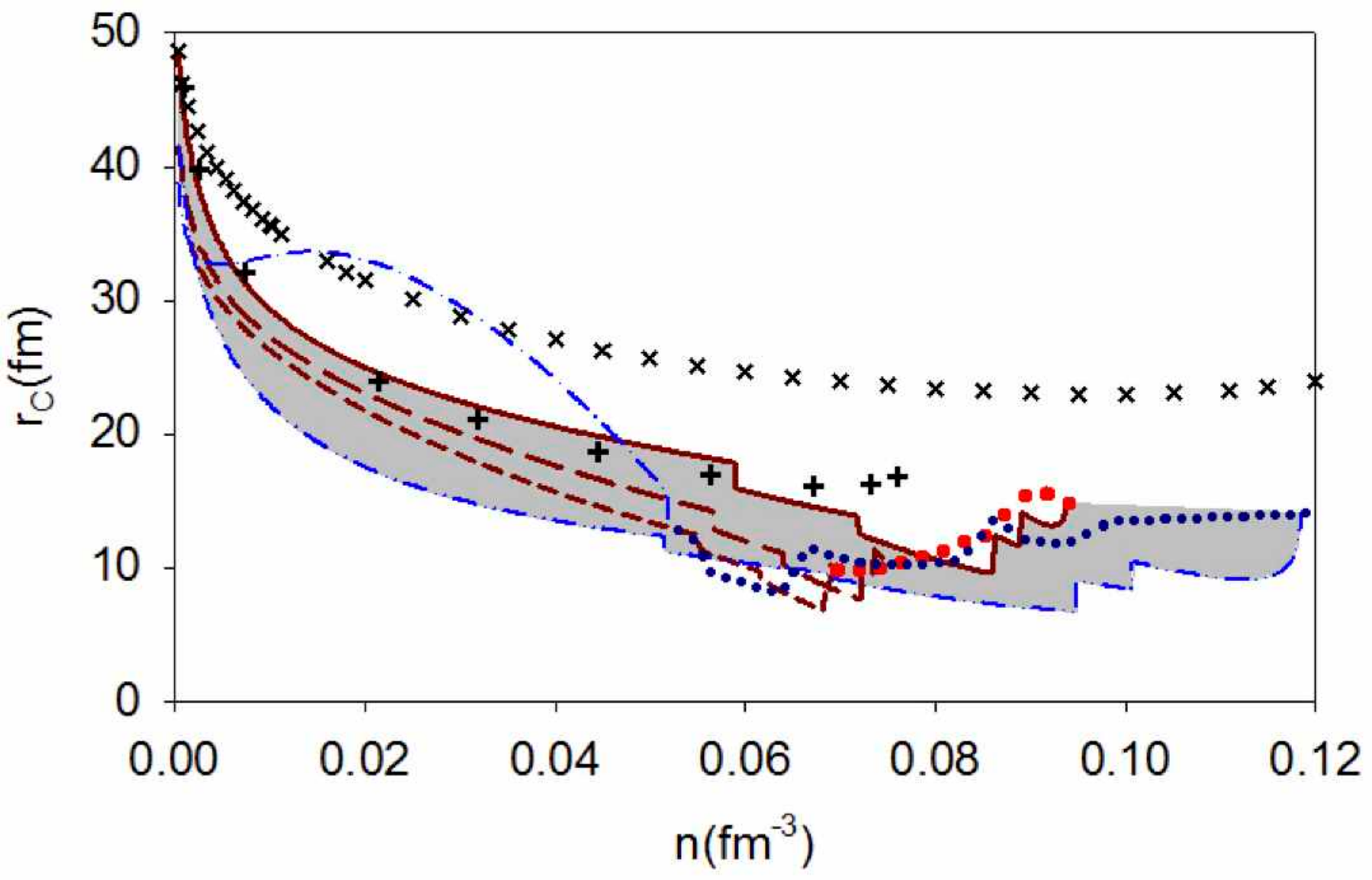}
\includegraphics[width=7cm,height=4.65cm,trim=15 0 0 0 ]{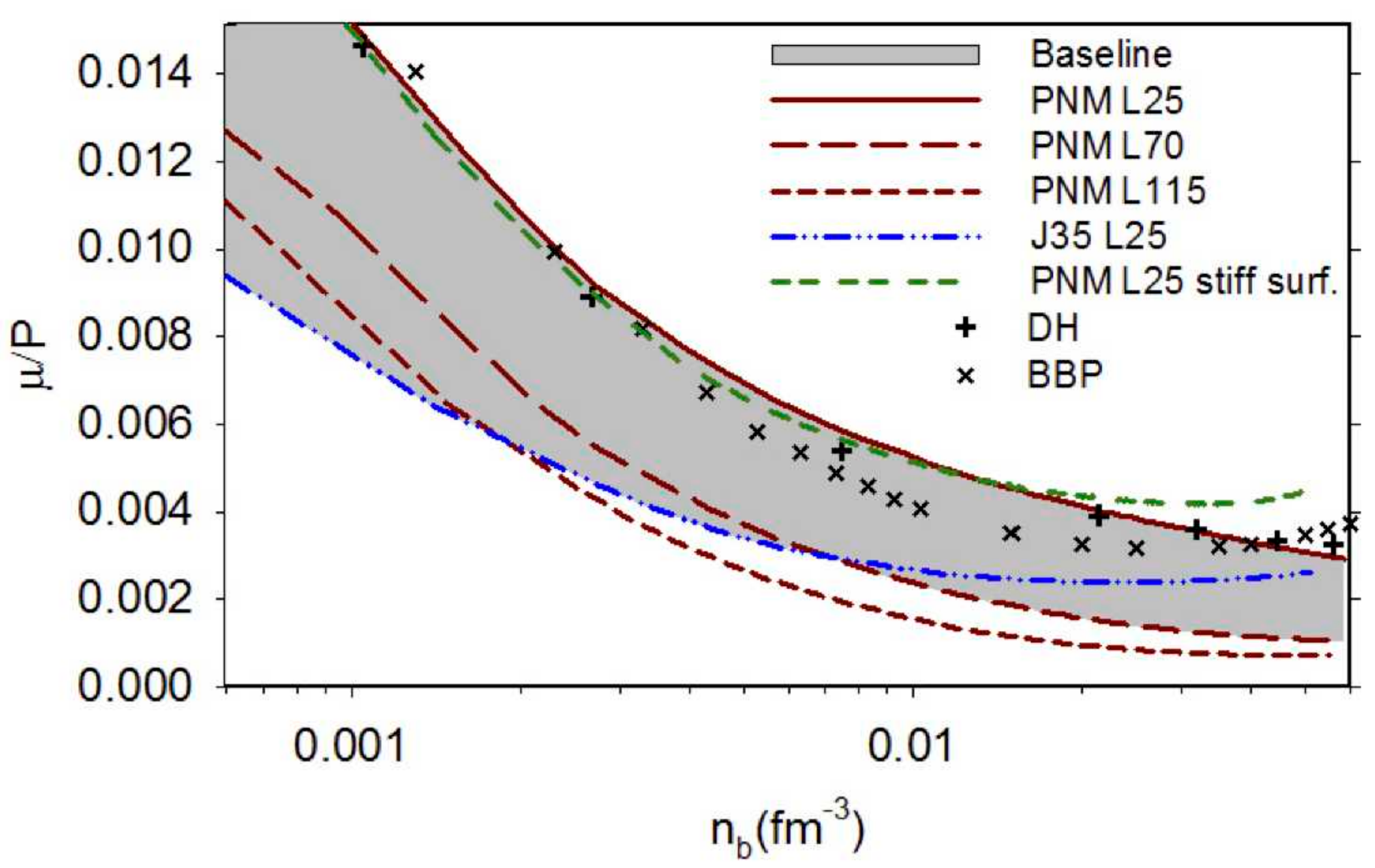}\includegraphics[width=7cm,height=4.65cm,trim=10 0 -10 0]{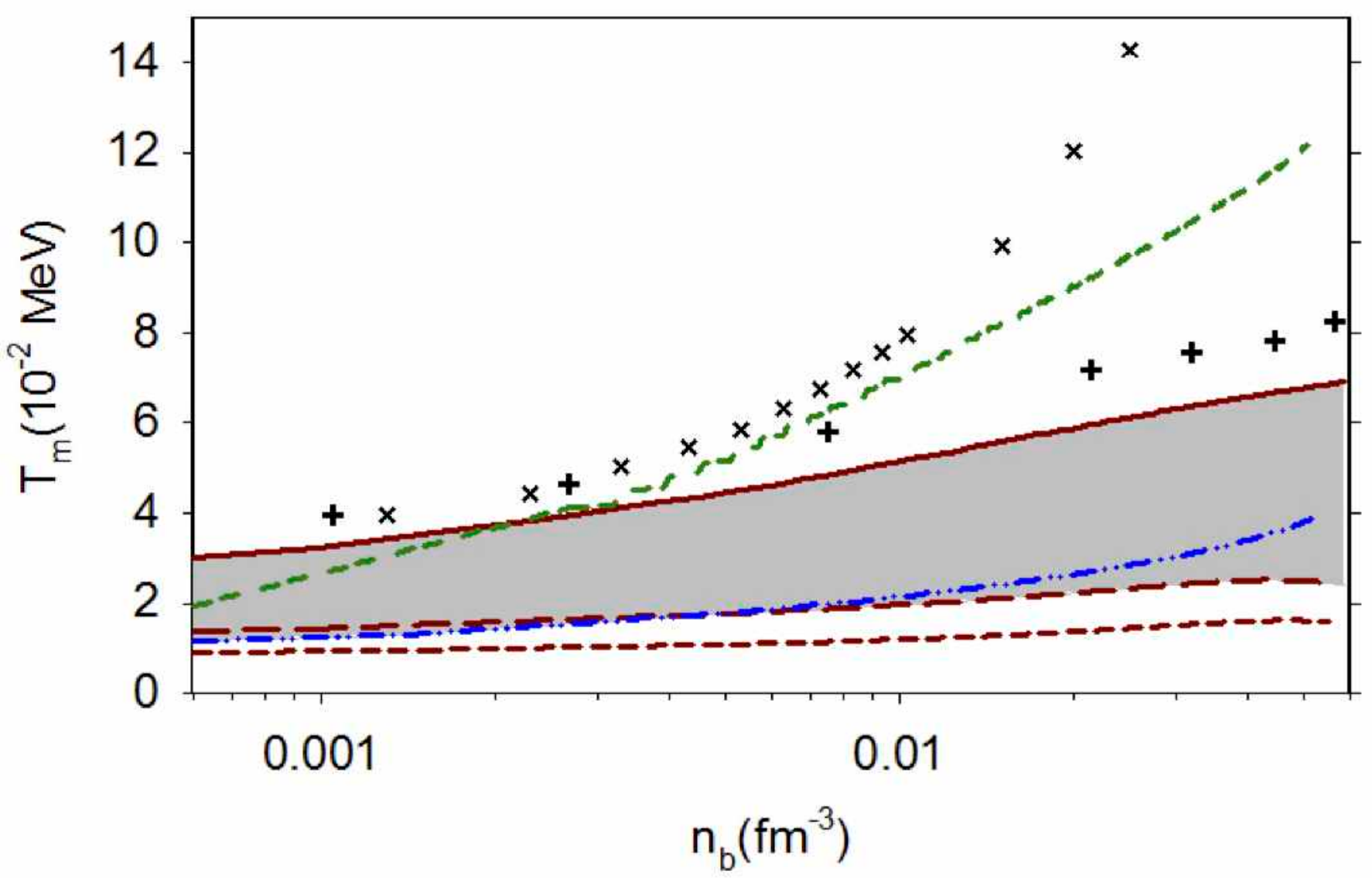}
\caption{ \textbf{Top left}: volume fraction $v$ occupied by nuclei/nuclear clusters, \textbf{top right}: density fraction $X_{\rm n}$ occupied by free neutrons, \textbf{middle left}: local proton fraction of clusters $x$, \textbf{middle right}: WS cell size $r_{\rm C}$, \textbf{bottom left}: shear modulus of crust scaled by pressure and \textbf{bottom right}: crustal melting temperature, versus baryon number density in the crust. Results are displayed for the $L=25, 70$ and 115 MeV members of the `J35' and `PNM' sequences. The loci of the transition densities over the range $L = 25 - 115$ MeV are displayed for the `PNM' sequence (larger black circles) and the `J35' sequence (smaller blue circles). The results from the BBP~\cite{BBP1971} and DH~\cite{Douchin2001} crust models are shown by the crosses and plusses respectively. The shaded region has the same meaning as in Fig.~4. The bottom two plots also show the results of the `PNM' $L$=25 MeV EOS with a stiff surface energy. Adaped from from \cite{Newton2011}.}
\end{center}
\end{figure}

We turn our attention to the composition of the inner crust/mantle versus density for a selection of EOSs from the `PNM' and `J35' sequences. Specifically, we show in Fig.~6 the volume fraction of the charged nuclear component $v$, the density fraction of dripped neutrons $X_{\rm n} = (1-v)n_{\rm n}/n_{\rm b}$, the local proton fraction of the charged nuclear component $x$, the WS cell size $r_{\rm C}$ and two properties of crustal matter estimated from the compositional parameters: the shear modulus $\mu$ \cite{Ogata1990, Strohmayer1991, Chugunov2010} at zero temperature and the melting temperature $T_{\rm m}$ (at which the crystalline lattice melts into a plasma) \cite{Chamel2008}:
\be \label{eq:shear_mod}
	\mu = 0.1106 \left(\frac{4\pi}{3}\right)^{1/3} A^{-4/3} n_{\rm b}^{4/3} (1-X_{\rm n})^{4/3} (Ze)^2; \;\;\;\;\; T_{\rm m} = { (Ze)^2 \over 175 k_{\rm B} r_{\rm C}}.
\ee

The volume fraction rises with density in the inner crust as the nuclei get larger and more closely spaced. The predictions of the various EOSs tend to converge close to the transition to the mantle, where the volume fraction approaches the $\sim0.125$ estimate of the the Bohr-Wheeler fission instability criterion (see, e.g. \cite{PethickRev1995}), shown here with the dashed line. Below this density, a higher $L$ corresponds to lower volume fractions: the correspondingly higher dripped neutron pressure favors smaller nuclei. Above the mantle transition, predictions vary widely depending on $L$ and the $L-J$ correlation. The volume fraction reaches $\textgreater 0.8$ at the crust-core transition within the baseline region, a warning that the WS approximation is certainly not valid at the highest crustal densities. The crust-core transition fraction remains high for the `PNM' sequence, but falls with increasing $L$ for the `J35' sequence down to 0.1 ($L$=115 MeV).

The density fraction of dripped neutrons increases sharply with density at the top of the inner crust, before leveling off at intermediate densities to $X_{\rm n} \approx 0.7-0.9$ with the higher values predictions of higher $L$.  Close to the inner-crust-mantle transition, $X_{\rm n}$ falls rapidly down to 0.02-0.2 at the crust-core transition, mirroring the behavior of $v$.

The local proton fraction $x$ generally decreases increasing density over the whole density range. For $J=35$ MeV, the crust-core transition fraction varies by about 0.03; from $L=25$ MeV, $x_{\rm cc}$ decreases from 0.06 down to 0.03 at $L=70$ MeV, then increasing back up to 0.06 at $L=115$ MeV; for PNM the variation is similar, starting at $x_{\rm cc}$=0.025 for  $L=25$ MeV and increasing with $L$ up to 0.06 at $L=115$ MeV. In the lower density region the variation in $x$ remains approximately constant at around 0.05. At a given density, higher $J$ and higher $L$ correlate with higher $x$. Higher $J$ favors a higher proton fraction; higher $L$ favors smaller, denser nuclei and nuclear clusters; as the symmetry energy increases with density, denser nuclei will also tend to favor higher $x$.

The WS cell size $r_{\rm C}$ decreases smoothly up to the transition to pasta, and then proceeds through a series of discontinuous jumps as matter transitions through the various nuclear shapes. In reality, these jumps may be smoothed out by the existence of intermediate shapes not considered in this work \cite{Nakazato2009}. Higher $L$, (higher neutron pressure) leads to smaller nuclei and smaller values of $r_{\rm C}$.

For comparison, we also show the results of two of the most widely used CLDM crust EOSs, from Baym, Bethe and Pethick (BBP) \cite{BBP1971} and Douchin and Haensel (DH) \cite{Douchin2001}. BBP over-predicts the strength of the surface energy of nuclei \cite{RBP1972}; DH calculates the surface energy consistently with the bulk energy using the SLy4 Skyrme parameterization ($L$=45.5 MeV, $J$=31.8 MeV). The DH EOS contains no pasta, a result of a slightly stiffer surface energy than in our baseline models (see discussion in \cite{Newton2011}).

The shear modulus is shown scaled to the crustal pressure and the melting temperature is given in terms of 0.01 MeV ($1.16 \times 10^8$ K). The baseline results for the shear modulus show a variation of a factor of about 2 throughout the inner crust; the melting temperature varies by a factor of 3 throughout the crust. The BBP and DH predictions are shown for comparison. The results from the $L$=25 MeV member of the `PNM' sequence is also shown with a stiffer surface energy, which elevates the melting temperature by up to a factor of 2 at high densities, but doesn't appreciably affect the baseline range for the shear modulus.

\section{Global crust properties}

An estimate of the mass and moment of inertia fractions of the mantle and the component of mantle containing the bubble phases (where the protons are delocalized) relative to the total \emph{crust} amounts can be obtained using ${\Delta M_{\rm i} / \Delta M_{\rm crust}} \approx 1 - {P_{\rm i} / {P_{\rm cc}}}$ (see \cite{LorenzPethick1993}) where

\begin{wrapfigure}{l}{7cm}\label{fig:7}
\vspace{-15pt}
\begin{center}
\includegraphics[width=7cm,height=5.5cm]{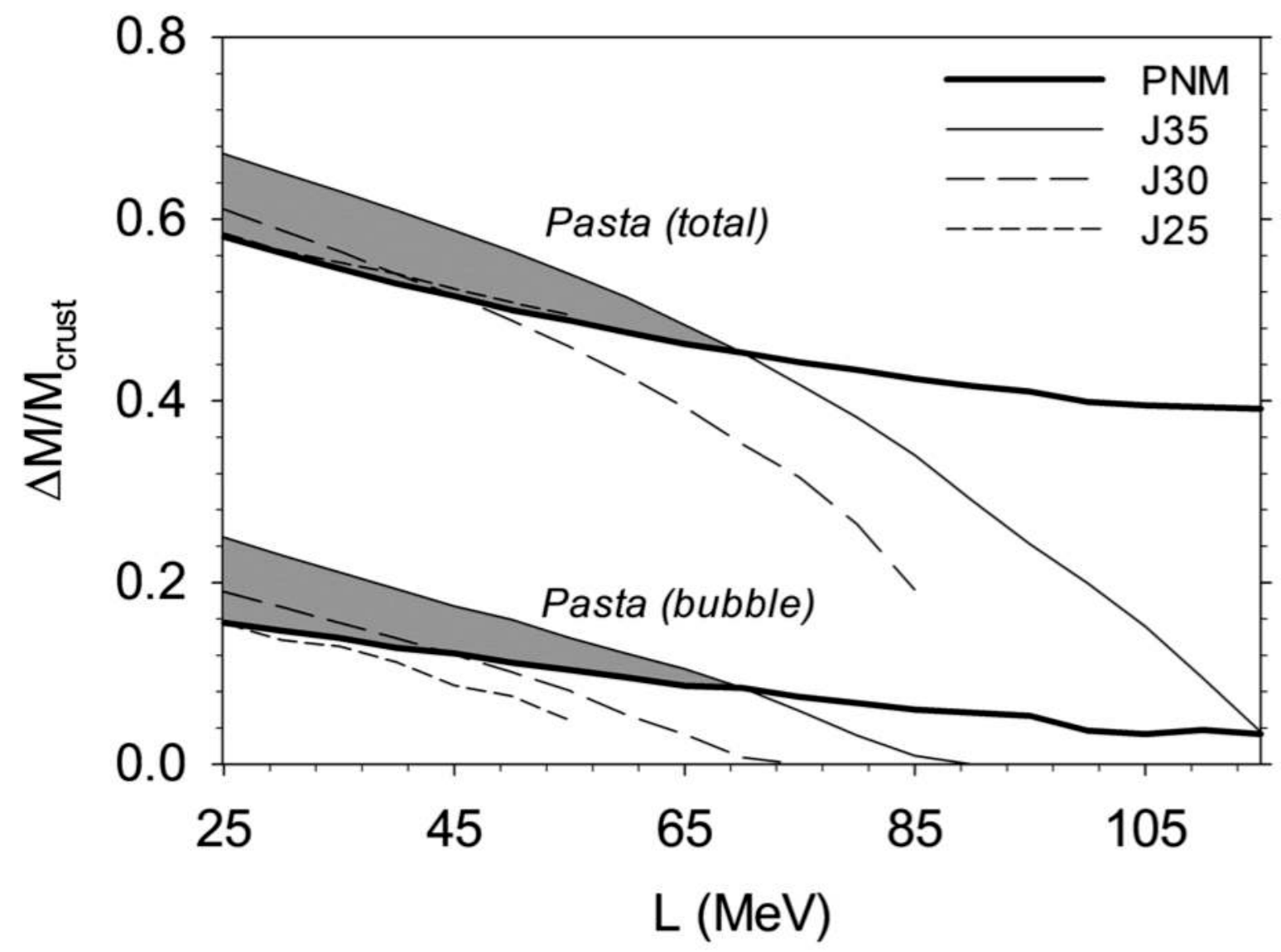}
\caption{Fraction of crustal mass occupied by the mantle and the bubble phases therein for the `PNM' sequence and three constant-$J$ sequences of EOSs, with the baseline region shaded. Taken from \cite{Newton2011}.}
\end{center}
\vspace{-15pt}
\end{wrapfigure}

\noindent ${\Delta M_i}$ is the mass and $P_i$ the pressure at the upper boundary of the crustal component $i=$ pasta, bubble phases.
The expressions for the moment of inertia fractions are identical. The mass fractions are shown in Fig.~7. For $L<70$ MeV, the pasta phases account for between 50\% and 70\% the mass and moment of inertia of the crust, a result of those phases existing in the highest density region of the crust. For $L>70$ MeV, the fractions remain above 40\% for the `PNM' sequence but fall off rapidly to just a few percent at $L=$ 115 MeV for the `J35' sequence.

The equivalent fractions for the bubble phases can also be significant: 10\% - 25\% for $L<70$ MeV, remaining above 10 \% for the `PNM' sequence and dropping to zero at $L\approx85$ MeV for `J35'. This region could therefore have an important influence on a range of phenomena, e.g. allowing the direct Urca process \cite{Gusakov2004}.

We extend our CLDM crust equations of state to the core using the same MSL EOS to calculate the pressure and energy density of beta-equilibrated nuclear matter in the core. At the highest densities, where a description of matter in terms of nucleonic degrees of freedom is expected to break down, we smoothly join the MSL EOS to two polytropic EOSs of the form $P = K \epsilon^{(1+1/n)}$ in a similar way to \cite{Steiner2010}. This also ensures that our complete EOS is always sufficiently stiff at the highest densities to produce 2$M_{\odot}$ neutron stars as demanded by observations \cite{Demo10}. The joins are made at energy densities of 300 MeV fm$^{-3}$ and 600 MeV fm$^{-3}$ by adjusting the constant $K$ to keep the pressure continuous at the join. The lower density polytrope has an index set at $n=0.5$, while the second index takes  a range $n=0.5-1.5$ for values of $L$ from 25-115 MeV respectively \cite{Wen11}. Note that although the additions of the polytropes substantially changes the maximum neutron star mass for small (soft) values of $L$, it does not substantially affect the radius and crust thickness of a neutron star of a given mass. Then, using the transition densities and pressures calculated in the CLDM model we can solve the general relativistic of hydrodynamic equilibrium (TOV) equations to obtain model static neutron stars and examine the global crust properties.

\begin{figure}[!tb]\label{fig:8}
\begin{center}
\includegraphics[width=7cm,height=6cm]{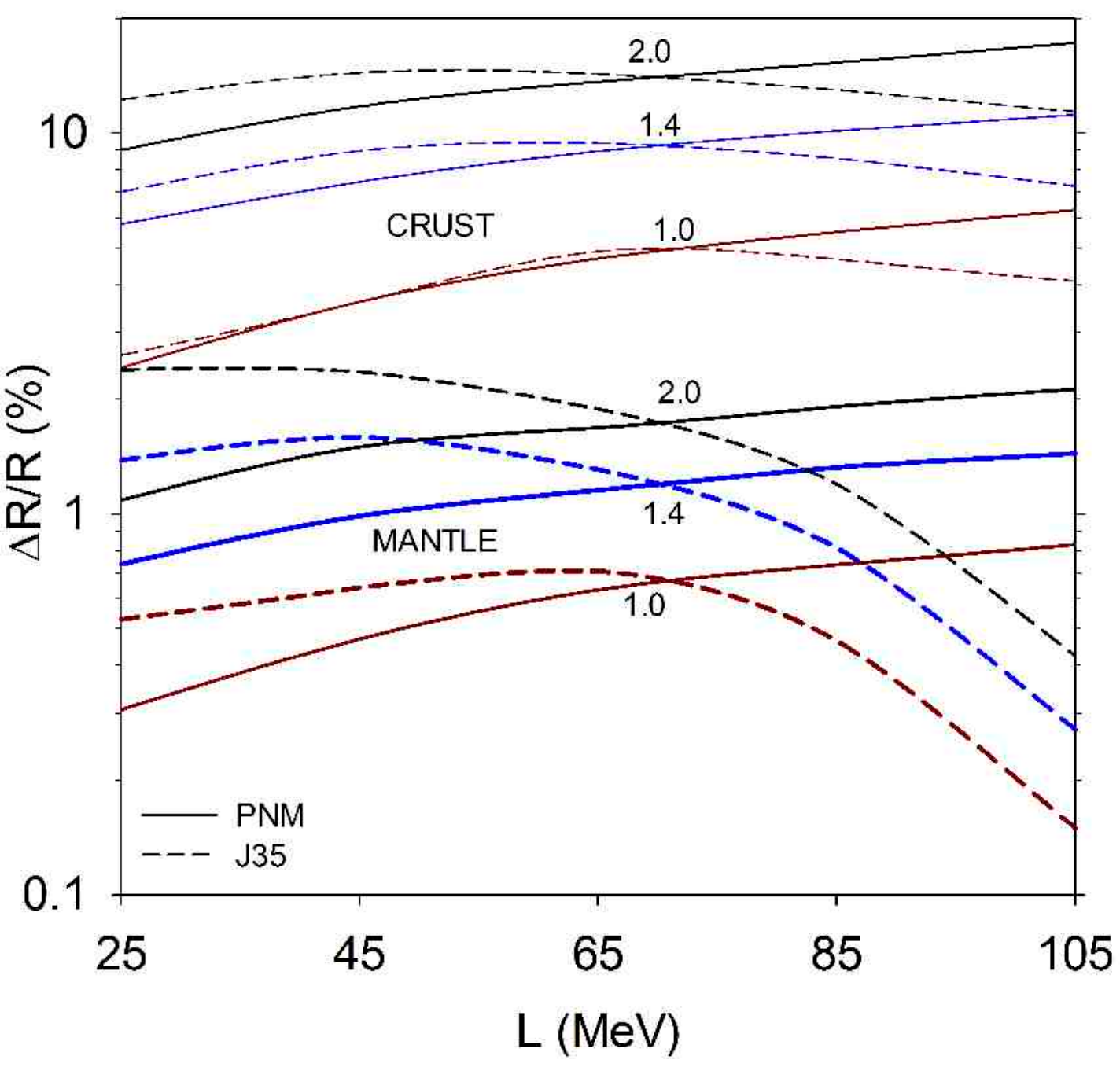}\includegraphics[width=7cm,height=6cm]{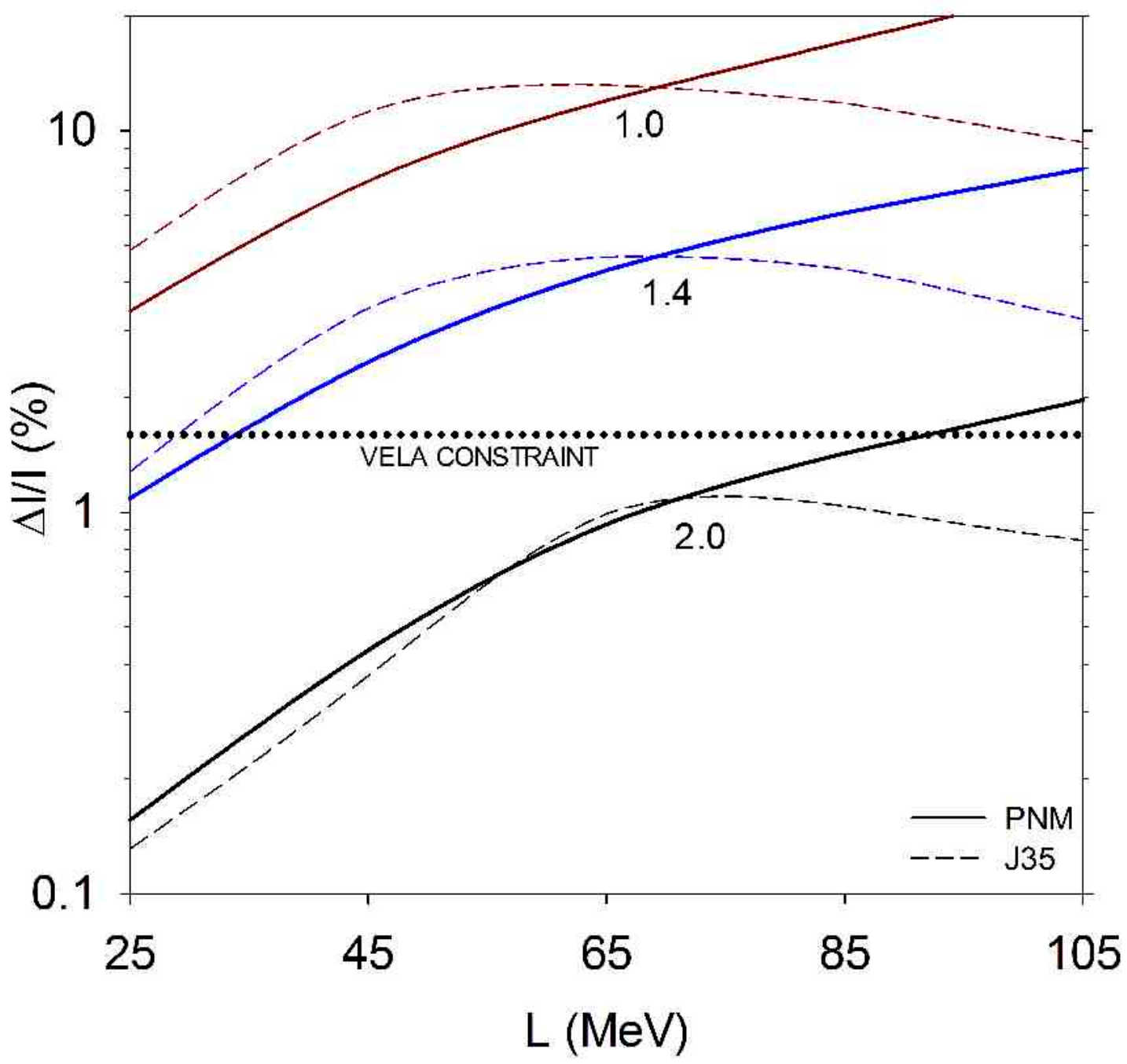}
\caption{Total crust and mantle thickness (\textbf{left}) and crustal moment of inertia (\textbf{right}) as a percentage of the total stellar radius and moment of inertia respectively. Results for the `PNM' and `J35' sequences are shown versus $L$ for 3 stellar masses: 1.0, 1.4 and 2.0 $M_{\odot}$. Taken from \cite{FuturePaper}.}
\end{center}
\end{figure}

Fig.~8 shows the percentage thickness of the whole crust and mantle and the crustal moment of inertia compared to the equivalent global quantities. The trends of these quantities with $L$ very much depend on the $J-L$ correlation: for the `PNM' sequence, the thicknesses and crustal moments of inertia rise monotonically with $L$, whereas for the `J35' sequence, the relationships are distinguished by a rise with $L$ up to a maximum at $L\approx 45-70$ MeV depending on mass, and then a shallower decline with $L$ beyond the maximum. The thicknesses change by a factor of $\sim 2$ over the range of $L$ and by an order of magnitude from 1.0 - 2.0 $M_{\odot}$; the mantle thickness is consistently about an order of magnitude lower than the total crust thickness. The relative crustal moment of inertia $\Delta I/I$ is a quantity of relevance in, e.g., the study of pulsar glitches. Under the assumption that glitches are self-regulating phenomena involving a constant time-averaged angular momentum transfer between some crustal component such as the dripped superfluid neutrons and the rest of the star, limits can be set on the minimum value of $\Delta I/I$ for a particular pulsar given sufficient observational data \cite{Espinoza2011,Link1999}. Current observational data on the Vela pulsar provides a limit of $\Delta I/I \ge 1.6\%$ \cite{Espinoza2011}, which is shown as the horizontal dashed line on the right plot. Taking the above assumptions at face value, one can deduce from the `PNM' sequence that stiffer EOSs are consistent with the Vela data for a wider range of masses; the softest EOS $L$=25 MeV is consistent only for $M \lesssim$1.25$M_{\odot}$. However, for the sequence `J35', the $L\textgreater85$ MeV EOSs are inconsistent with the Vela constraint as are the $L\textless50$ MeV EOSs for $M\approx$1.7$M_{\odot}$. It should be emphasized that in determining the crustal thickness and moment of inertia, $L$ alone is insufficient; they depend sensitively on how $L$ correlates with $J$.

\section{Dependence of observable quantities on symmetry energy}

Finally we show two examples of simple estimates of potential observables that incorporate crust composition, thickness and global stellar properties discussed in the previous sections.

\begin{figure}[!tb]\label{fig:11}
\begin{center}
\includegraphics[width=7cm,height=9cm]{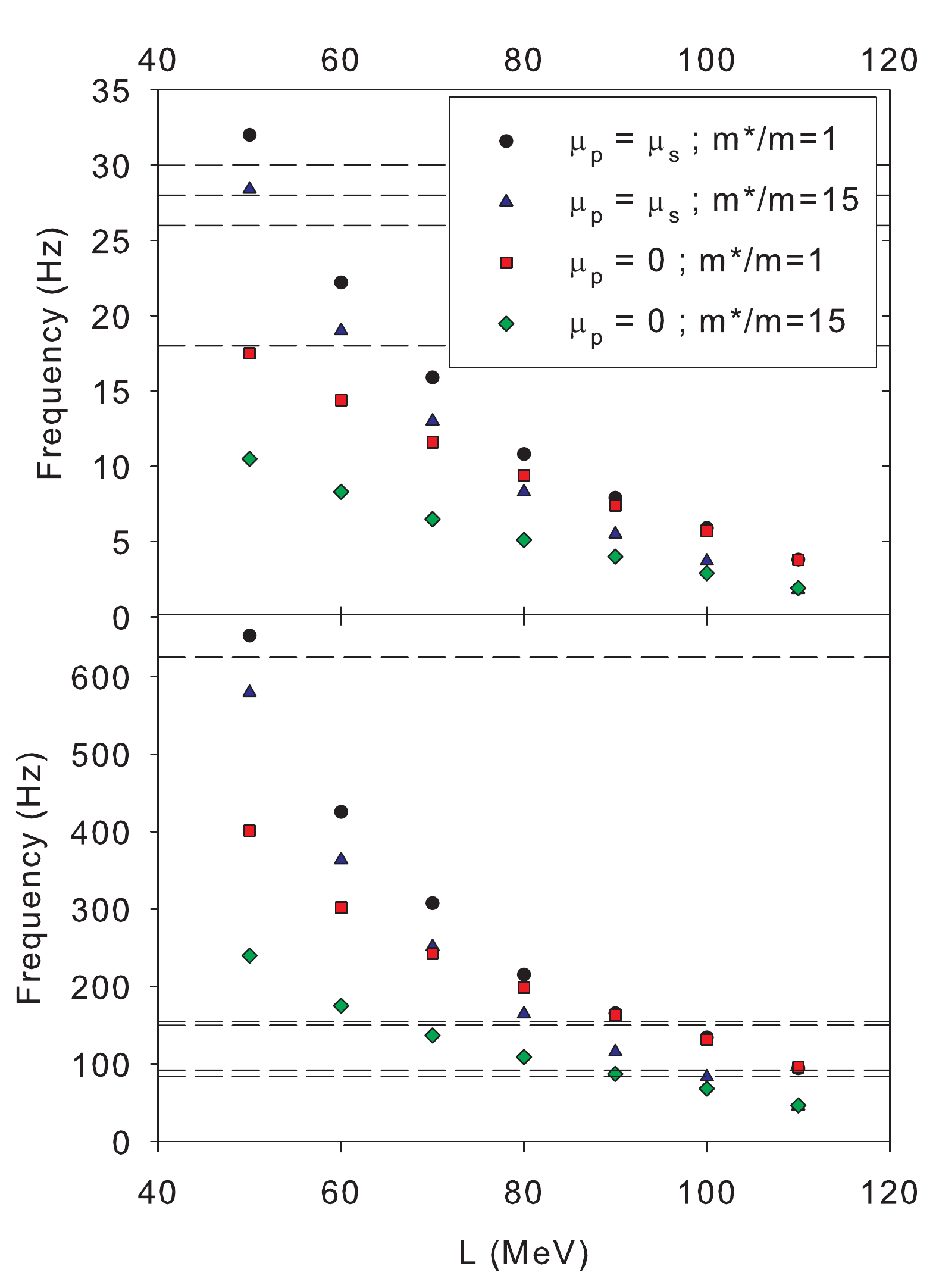}\includegraphics[width=7cm,height=9cm]{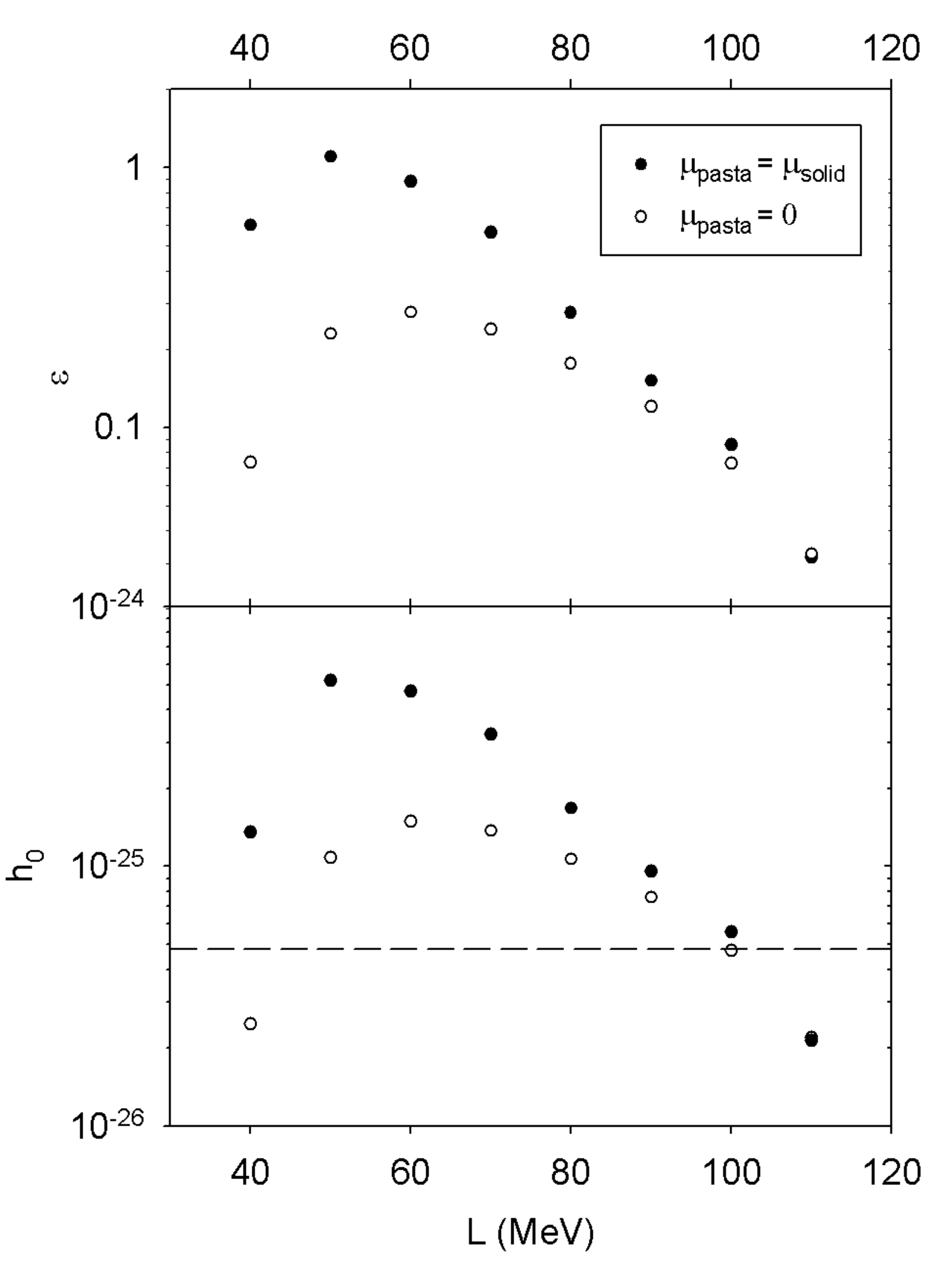}
\caption{\textbf{Left panel}: The frequency of the fundamental torsional oscillation mode (top) and the first overtone (bottom) in the the crust for a 1.4$M_{\odot}$ star as a function of $L$. The circles and triangles show the frequency assuming the shear modulus in the pasta phases to be that of a Coulomb lattice; the squares and diamonds show the frequency assuming the shear modulus in the pasta phase to be zero; the triangles and diamonds take into account the entrainment of the superfluid neutrons by the nuclear clusters. The dashed lines show candidate frequencies for the fundamental modes; 18, 26, 28, 30Hz in the fundamental frequency range and 84, 92, 150, 155Hz in the range of the first overtone. \textbf{Right panel}: Estimate of the maximum quadrupole ellipticity of a 1.4$M_{\odot}$ neutron star (top) and the corresponding gravitational wave strain (bottom) as a function of $L$. The strain is calculated assuming a neutron star frequency of 300Hz and a distance of 0.1kpc from Earth. The filled circles indicate the value taking the shear modulus in the pasta phases to be that of a solid Coulomb lattice; the empty circles indicate the value taking the shear modulus in the pasta phase to be zero. The dashed line in the bottom plot indicates the sensitivity of the most recent LIGO science run \cite{Abbott2010}. Taken from \cite{Gearheart2011}.}
\end{center}
\end{figure}

The left panel of Fig.~9 shows the fundamental frequency $l$=2, $n$=0 and first overtone $l$=2, $n$=1 from the spectrum of torsional shear oscillations in the crust for a $1.4 M_{\odot}$ star ($n$ is the number of radial nodes and $l$ the angular constant associated with the spherical harmonics $Y_l^m$) \cite{Gearheart2011}. Observed values of the frequency of QPOs at 16, 26, 28, 30, 84, 92, 150, 155 and 625 Hz \cite{Israel2005, Watts2006, Strohmayer2005, Strohmayer2006} are indicated by the dashed lines. The calculation uses the value of the shear modulus at either the crust-core boundary, assuming the mantle is an elastic solid, or the inner crust-mantle boundary, assuming the mantle is a liquid (an upper limit on the difference in mechanical properties of the pasta from the rest of the crust). The `PNM' sequence of EOSs is used. Superfluid effects are estimated through the relative mesoscopic neutron effective mass $m^*/m$ \cite{Chamel2005b}; a value of 1 corresponds to no superfluid entrainment effects. The frequency generically decreases with increasing $L$. Ignoring superfluid effects and the effects of the pasta phases, the frequency matches observed QPOs from SGRs only at the lowest values of $L$. If the pasta phases are purely liquid, the frequency falls by a factor of 3, making it difficult to match the 28Hz frequency observed, and being consistent with the 18Hz observed frequency only at the lowest value of $L$. One can see that, accepting the model and the interpretations of the results at face value, observed fundamental frequencies only match the predictions for $L<70$ MeV.

The right panel of Fig.~11 shows the maximum quadrupole ellipticity of the crust of a $1.4 M_{\odot}$ star, normalized to the canonical value of $10^{-7} (\sigma_{\rm break}/10^{-2})$, as a function of $L$ for the same `PNM' sequence of EOSs. The equivalent gravitational wave strain is plotted in the bottom half for an `ideal' reference neutron star with frequency 300Hz and a distance of 0.1kpc. The convolution of the trends of global star and crust properties and crust composition with $L$ manifests itself in the non-monotonic variation of ellipticity with $L$, having a maximum at $L\approx$ 50-60 MeV. The sensitivity of the most recent LIGO science run is indicated by the dashed line; for this ideal neutron star, the GW strain from crustal mountains predicted in this simple model is detectable for $L\approx$ 40-100 MeV offering the chance to distinguish between the various possibilities of crust properties. 

\section{Conclusions and future directions}

We have constructed sets of neutron star EOSs that consistently encompass the inner crust and core, and include the crust composition and transition densities, using the compressible liquid drop model for the crust. These have been used to demonstrate the effect of the magnitude $J$ and slope $L$  of the symmetry energy at nuclear saturation density on microscopic and global crustal properties, and potential neutron star observables. The crust-core transition density and pressure, crustal composition, stellar and crustal mass, thickness and moment of inertia, torsional crust oscillation frequencies and maximum crust deformation all depend sensitively on both $J$ and $L$ within their experimentally constrained ranges. One of the dominant neutron star model dependences is therefore the correlation between $J$ and $L$ which constrains their possible values in $J-L$ space. Experimental and theoretical information about $L$ and $J$ and their correlation will continue to improve; in order to add neutron star observations to this investigation, consistent explorations of neutron star properties over the constrained ranges is of great importance. The sets of EOSs used in this paper are available to interested parties \cite{mywebsite}.

The simplicity of the CLDM allows useful exploration of the dependence of composition and transition density on $J$ and $L$, but it possesses several drawbacks. Firstly shell effects within the nuclei, or arising from scattering of dripped neutrons off of nuclear clusters, are ignored. Such shell effects can dominate the determination of nuclear geometry, the equilibirium size of the nuclear clusters (which will proceed in discrete jumps corresponding to changes in the nuclear `magic' numbers with density) and the ordering of the pasta phases, as well as transport properties such as contributions to heat transport from nuclear components \cite{Sandulescu2007} and entrainment of dripped neutrons by clusters \cite{Chamel2005b}. Secondly, the WS approximation is expected to break down at when the nuclear separation becomes comparable to the cell size \cite{Chamel2007}, which occurs in the mantle. Thirdly, effects that act over ranges greater than the unit cell are not consistently accommodated in the CLDM; longer range electron screening, larger scale self-organization of pasta phases and long wavelength transport effects are all unaccounted for.


\begin{figure}[!t]\label{fig:12}
\begin{minipage}{6.6cm}
\begin{center}
\includegraphics[width=3.3cm,height=3cm]{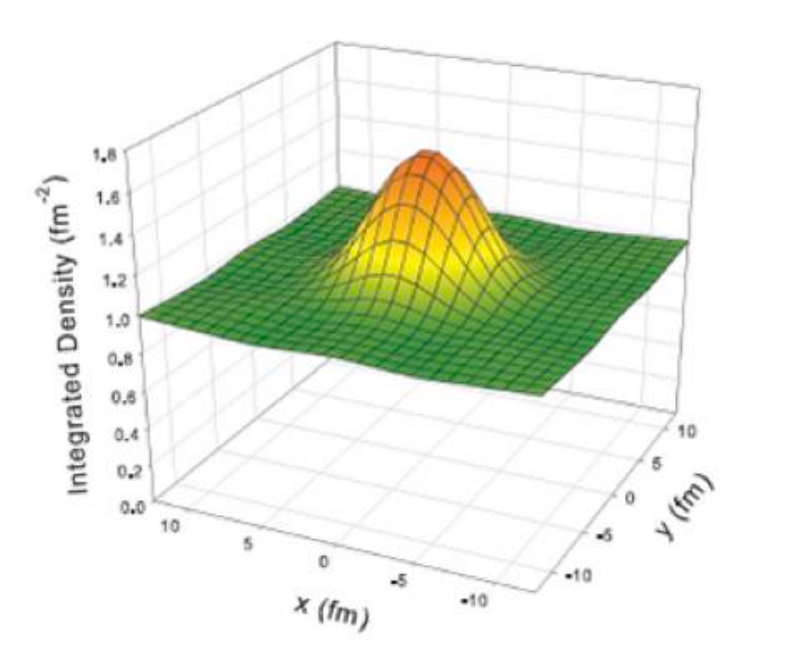}\includegraphics[width=3.3cm,height=3cm]{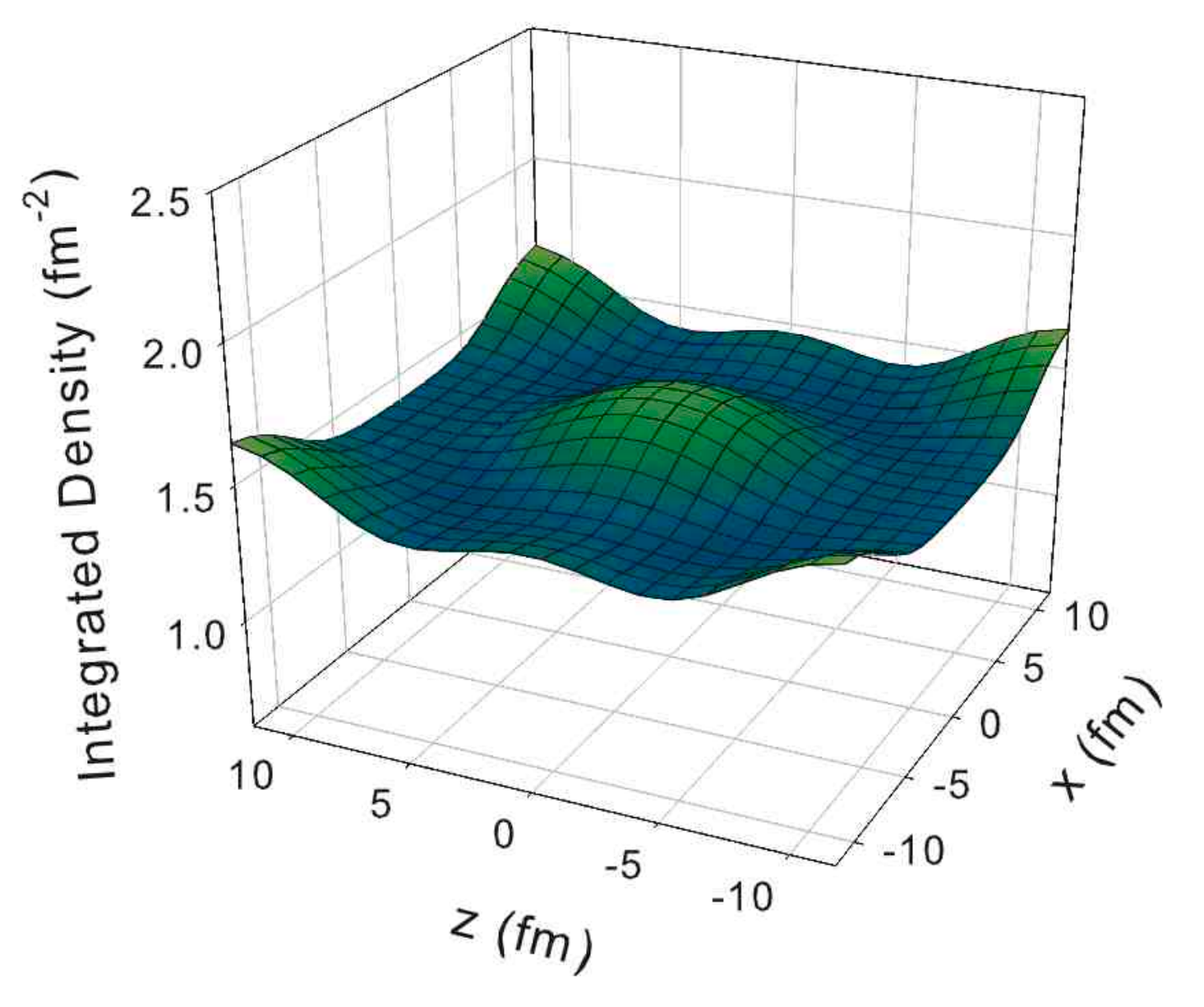}\\
\includegraphics[width=3.3cm,height=3cm]{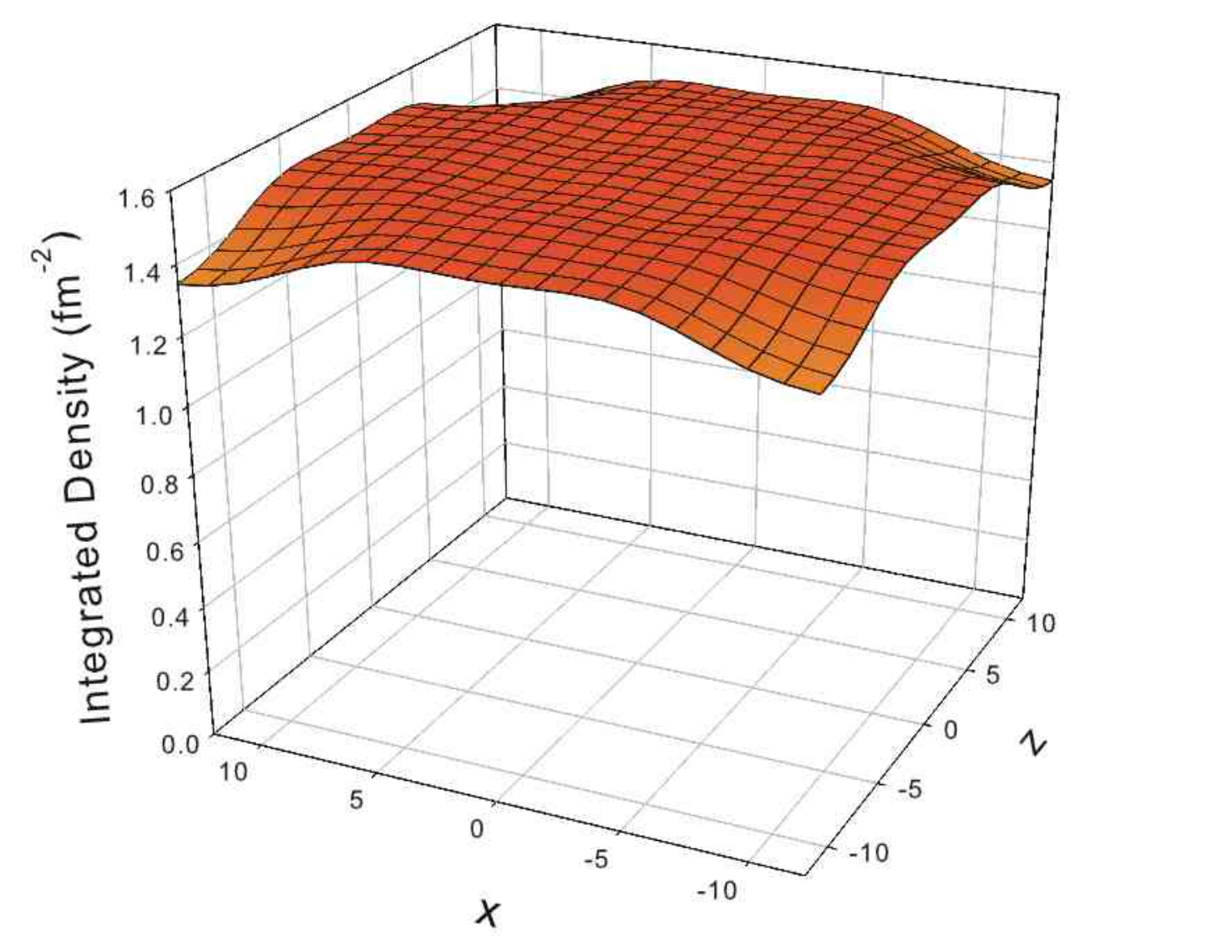}\includegraphics[width=3.3cm,height=3cm]{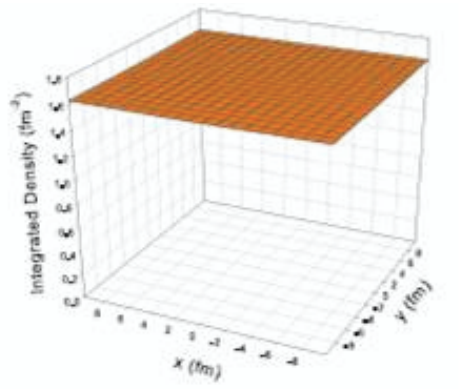}
\end{center}
\caption{Integrated neutron density in a cubic cell at densities of $n$=0.04, 0.06, 0.08 and 0.1 fm$^{-3}$ (top left to bottom right) calculated using the 3DHF method \cite{Newton2011} with the SLy4 Skyrme parameterization.}
\end{minipage}
\hfill
\begin{minipage}{6.6cm}
\begin{center}
\includegraphics[width=6.6cm,height=6cm]{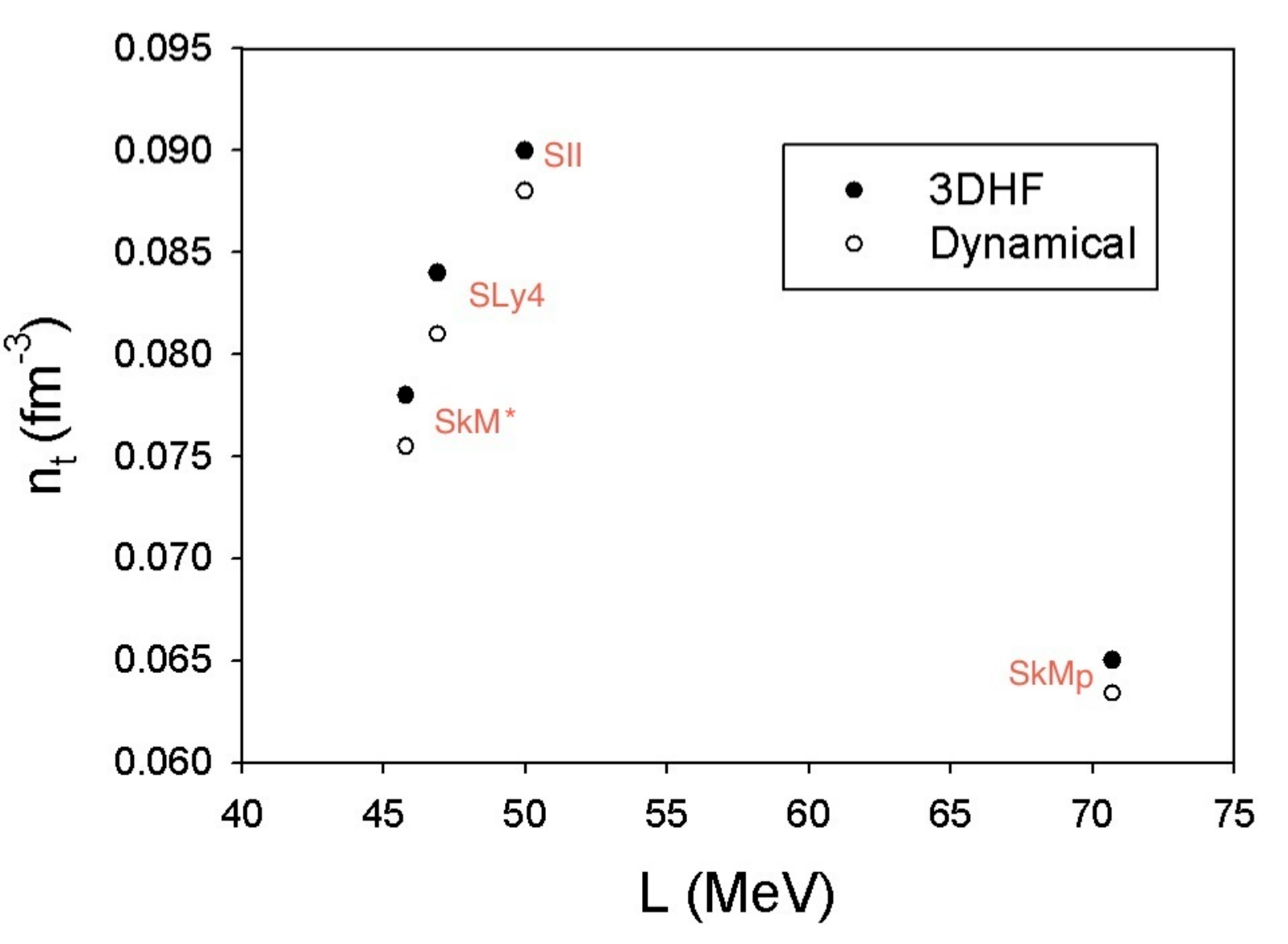}
\end{center}
\caption{Transition densities obtained using the 3DHF method compared to those obtained using the dynamical method of \cite{Xu2009} for the SLy4, SII, SkM* and SkMp Skyrme parameterizations.}
\end{minipage}
\vspace{10pt}
\end{figure}

Some of these effects can be taken into account by more sophisticated crust models. The Thomas-Fermi and Extended Thomas-Fermi methods (e.g \cite{Buchler1971,Oyamatsu1993,Cheng1997}) are semi-classical models employing the local density approximation which allows the nuclear surface energy to be calculated self-consistently with the bulk nuclear energy. Shell corrections can be self-consistently added using the Strutinsky integral method \cite{Onsi2008}. The 1D-Hartree-Fock method is fully microscopic and accounts for surface and shell energies self-consistently (e.g. \cite{Sandulescu2007,Negele1973,Montani2004,Baldo2005}), but it describes only spherically symmetric configurations and is thus constrained by the spherical Wigner-Seitz approximation. This restriction is lifted in the more computationally demanding 3D-Hartree-Fock (3DHF) method \cite{Magierski2002,Gogelein2008}, allowing a self-consistent probe of the shape-phase-space of pasta layers with shell effects included. As example, some results of a 3DHF model \cite{Newton2009} are shown in Figs.~10 and ~11. The former shows the local neutron density is plotted over the unit cell at four different densities encompassing the crust-core transition, showing the evolution of nuclear shape from spherical through bubble to uniform matter. The latter shows the resulting transition densities for four different parameterizations of the Skyrme interaction, compared to the dynamical method of \cite{Xu2009}. The Hartree-Fock method is naturally extended to include pairing effects self-consistently (Hartree-Fock-Bugoliubov). Longer range effects can be simulated via the semi-classical quantum molecular dynamics method \cite{Maruyama1998,Horowitz2004,Watanabe2001,Sonoda2007}. Between these models, a complete physical description of the crust can be built up. Many of the methods mentioned are much more time-consuming than the CLDM, making a wide-ranging survey of nuclear matter parameters unwieldy. However, if we know the quantities for which the CLDM provides a reasonable estimate, and to what densities it remains reasonable, one can use the CLDM as a useful guide for more realistic calculations. Ultimately, the goal should be to have all relevant microscopic inputs to neutron star models calculated consistently with the nuclear matter EOS; much work still needs to be done in this direction.

\section*{ACKNOWLEDGEMENTS}
This work is supported in part by the National Aeronautics and Space Administration under grant NNX11AC41G issued through the Science Mission Directorate and the National Science Foundation under grants PHY-0757839 and PHY-1068022 and the Texas Coordinating Board of Higher Education under grant No. 003565-0004-2007.

\label{lastpage-01}


\begin{thebibliography}{149}


\bibitem{BPS1971}
G. Baym, C.J.Pethick and P. Sutherland, ApJ \textbf{170}, 299 (1971).

\bibitem{BBP1971} 
G. Baym, H.A. Bethe and C.J. Pethick, Nucl. Phys. \textbf{A175}, 225 (1971).

\bibitem{Dean2003}
D.J. Dean and M. Hjorth-Jensen, Rev. Mod. Phys. \textbf{75}, 607, (2003).

\bibitem{Gusakov2004} 
M.E. Gusakov, A.D. Kaminker, O.Y. Gnedin and P. Haensel, A\&A \textbf{421}, 1143 (2004).

\bibitem{Ravenhall1983} 
D.G. Ravenhall, C.J. Pethick and J.R. Wilson, Phys. Rev. Lett. \textbf{50}, 26, 2066 (1983).

\bibitem{Oyamatsu1984}
K. Oyamatsu, M. Hashimoto and M. Yamada, Prog. Th. Phys. \textbf{72}, 2, 373 (1984).

\bibitem{Jones2002}
R.A.L. Jones,  \emph{Soft Condensed Matter}, OUP, Oxford, U.K. (2002)

\bibitem{Watanabe2005}
G. Watanabe and H. Sonoda, cond-mat/0502515 (2005).

\bibitem{Pethick1998}
C.J. Pethick and A.Y. Potekhin, Phys. Lett. \textbf{B427}, 7 (1998).

\bibitem{Douchin2001}
F. Douchin and P. Haensel, A\&A \textbf{380}, 151 (2001).

\bibitem{Lattimer2001}
J.M. Lattimer and M. Prakash, ApJ \textbf{550}, 1, 426 (2001).


\bibitem{Espinoza2011}
C.M. Espinoza, A.G. Lyne, B.W. Stappers and M. Kramer, MNRAS \textbf{414}, 167 (2011).

\bibitem{Baym1969}
G. Baym, C.J. Pethick and D. Pines, Nature \textbf{224}, 872 (1969).

\bibitem{Anderson1975}
P.W. Anderson and N. Itoh, Nature \textbf{256}, 25 (1975).

\bibitem{Link1999}
B. Link, R.I. Epstein and J.M. Lattimer, Phys. Rev. Lett. \textbf{83}, 17, 3362 (1999).

\bibitem{Ruderman1998}
M. Ruderman, T. Zh  and K. Chen, ApJ \textbf{492}, 267 (1998).


\bibitem{Stairs2000}
I.H. Stairs, A.G. Lyne, and S.L. Shemar, Nature \textbf{406}, 484 (2000).

\bibitem{Shabanova2001}
T.V. Shabanova, A.G. Lyne, and U.O. Urama, ApJ \textbf{552}, 321 (2001).

\bibitem{Cutler2002}
C. Cutler, Phys. Rev. \textbf{D66}, 084025 (2002).

\bibitem{Wasserman2003}
I. Wasserman, MNRAS \textbf{341}, 1020 (2003).

\bibitem{Cutler2003}
C. Cutler, G. Ushomirsky, and B. Link, ApJ \textbf{588}, 975 (2003).

\bibitem{Jones2001}
D.I. Jones and N. Andersson, MNRAS \textbf{324}, 811 (2001).

\bibitem{Link2003}
B. Link, Phys. Rev. Lett. \textbf{91}, 101101 (2003).

\bibitem{Link2006}
B. Link, A\&A \textbf{458}, 3, 881 (2006).

\bibitem{Glampedakis2009}
K. Glampedakis, N. Andersson and D.I. Jones, MNRAS \textbf{394}, 1908 (2009).


\bibitem{Israel2005}
G. Israel \emph{et al},  ApJ \textbf{628}, L53 (2005).

\bibitem{Watts2006}
A.L. Watts and T.E. Strohmayer, ApJ, \textbf{637}, L117 (2006).

\bibitem{Strohmayer2005}
T.E. Strohmayer and A.L. Watts, ApJ \textbf{632}, L11 (2005).

\bibitem{Strohmayer2006}
T.E. Strohmayer and A.L. Watts, ApJ \textbf{653}, 593 (2006).

\bibitem{SteinerWatts2009}
A.W. Steiner and A.L. Watts, Phys. Rev. Lett. \textbf{103}, 181101 (2009).

\bibitem{Samuelsson2007}
L. Samuelsson and N. Andersson, MNRAS \textbf{374}, 256 (2007).

\bibitem{Andersson2009}
N. Andersson, K. Glampedakis and L. Samuelsson, MNRAS \textbf{396}, 894 (2009).

\bibitem{Gearheart2011}
M. Gearheart, W.G. Newton, J. Hooker and Bao-An Li, accepted for publication in MNRAS (2011), arxiv:1106.4875

\bibitem{Sotani2011}
H. Sotani, MNRAS \textbf{417}, 1, L70 (2011).


\bibitem{Lattimer1994}
J.M. Lattimer, K.A. van Riper and M. Prakash, ApJ \textbf{425}, 802 (1994).

\bibitem{Gnedin2001}
O.Y. Gnedin, D.G. Yakovlev, and A.Y. Potekhin, MNRAS \textbf{324}, 725 (2001)


\bibitem{Andersson2011}
N. Andersson, V. Ferrari, D.I. Jones, K.D. Kokkotas, B. Krishnan, J.S. Read, L. Rezzolla and B. Zink, Gen. Rel. Grav. \textbf{43}, 2, 409 (2011).

\bibitem{Abbott2010}
B.P. Abbott, ApJ \textbf{713}, 671 (2010).


\bibitem{Bildsten1998}
L. Bildsten, ApJ \textbf{501}, L89 (1998).

\bibitem{Andersson1999}
N. Andersson, K.D. Kokkotas and N. Stergioulas, ApJ \textbf{516}, 307 (1999).

\bibitem{Bildsten00}
L. Bildsten and G. Ushomirsky, ApJ \textbf{529}, L33(2000).

\bibitem{Andersson00}
N. Andersson, D. I. Jones, K. D. Kokkotas, and N. Stergioulas, ApJ \textbf{534}, L75 (2000).

\bibitem{Lindblom00}
L. Lindblom, B.J. Owen, and G. Ushomirsky, Phys. Rev. D  \textbf{62}, 084030(2000).

\bibitem{Rieutord01}
M. Rieutord,  ApJ \textbf{550}, 443(2001).

\bibitem{Peralta06}
C. Peralta, A. Melatos, M.  Giacobello, and A. Ooi,  ApJ \textbf{644}, L53 (2006).

\bibitem{Glampedakis06}
K. Glampedakis, and N. Andersson, Phys. Rev. D \textbf{74}, 044040 (2006).

\bibitem{Wen11}
De-Hua Wen, W.G. Newton and Bao-An Li, submitted to Phys. Rev. C; arxiv:1110.5985 (2011).


\bibitem{Ushomirsky2000}
G. Ushomirsky, C. Cutler, L. Bildsten, MNRAS \textbf{319}, 902 (2000).

\bibitem{Haskell2006}
B. Haskell, D.I. Jones, N. Andersson, MNRAS \textbf{373}, 1423 (2006).


\bibitem{MSL01}
Lie-Wen Chen, Bao-Jun Cai, Che Mong Ko, Bao-An Li, Chun Shen and Jun Xu, Phys. Rev. \textbf{C80}, 014322 (2009)

\bibitem{LieWenChen2010} 
Lie-Wen Chen, Che Ming Ko, Bao-An Li and Jun Xu, Phys. Rev. \textbf{C82}, 024321 (2010).


\bibitem{Tsang2004}
M.B. Tsang et al., Phys. Rev. Lett. \textbf{92}, 062701 (2004).

\bibitem{Tsang2009} 
M.B. Tsang, Yingzun Zhang, P. Danielewicz, M. Famiano, Zhuxia Li, W.G. Lynch and A.W. Steiner, Phys. Rev. Lett. \textbf{102}, 122701 (2009).

\bibitem{Famiano2006} 
M.A. Famiano et al., Phys. Rev. Lett. \textbf{97}, 052701 (2006).

\bibitem{Chen2005} 
Lie-Wen Chen, Che Ming Ko and Bao-An Li, Phys. Rev. Lett. \textbf{94}, 032701 (2005).

\bibitem{Li2005} 
Bao-An Li and Lie-Wen Chen, Phys. Rev. \textbf{C72}, 064611 (2005).

\bibitem{Shetty2007} 
D.V. Shetty, S.J. Yennello, G.A. Souliotis, Phys. Rev. \textbf{C76}, 024606 (2007).

\bibitem{Danielewicz2009} 
P. Danielewicz and J. Lee, Nucl. Phys. \textbf{A818}, 1-2, 36, 818 (2009).

\bibitem{Centelles2009} 
M. Centelles, X. Roca-Maza, X. Vinas and M. Warda, PRL \textbf{102}, 122502 (2009).

\bibitem{Warda2009} 
M. Warda, X. Vinas, X. Roca-Maza, and M. Centelles, Phys. Rev. \textbf{C80}, 024316 (2009).

\bibitem{Klimkiewicz2007} 
A. Klimkiewicz \emph{et al}, Phys. Rev. \textbf{C76}, 051603(R) (2007).

\bibitem{Carbone2010} 
A. Carbone et al., Phys. Rev. \textbf{C81}, 041301 (R) (2010).

\bibitem{ChangXu2010} 
Chang Xu, Bao-An Li and Lie-Wen Chen, Phys. Rev. \textbf{C82}, 054607 (2010).


\bibitem{Liu2010}
M. Liu, N. Wang, Z.-X. Li and F.-S. Zhang, Phys. Rev. \textbf{C82}, 064306 (2010).

\bibitem{Oyamatsu2010}
K. Oyamatsu and K. Iida, Phys. Rev. \textbf{C81}, 054302 (2010).

\bibitem{Hebeler2010}
K. Hebeler and A. Schwenk, Phys. Rev. \textbf{C82}, 014314 (2010).

\bibitem{Gezerlis2008}
A. Gezerlis and J. Carlson, Phys. Rev. \textbf{C77}, 032801 (2008).

\bibitem{Gandolfi2011}
S. Gandolfi, J. Carlson and S. Reddy, arXiv:1101.1921.

\bibitem{FSUGold}
B. G. Todd-Rutel and J. Piekarewicz, Phys. Rev. Lett. 95, 122501 (2005);

\bibitem{Fattoyev2010}
F.J. Fattoyev and J. Piekarewicz, Phys. Rev. \textbf{C82} 025810 (2010).

\bibitem{Moustakidis2010}
Ch.C. Moustakidis, T. Niksic, G.A. Lalazissis, D. Vretenar and P. Ring, Phys. Rev. \textbf{C81}, 065803 (2010).

\bibitem{Ducoin2011}
C. Ducoin, J. Margueron, C. Providencia and I. Vidana, Phys. Rev. \textbf{C83}, 4, 045810 (2011).

\bibitem{Newton2011}
W.G. Newton, M. Gearheart and Bao-An Li, submitted to Phys. Rev. C; arxiv:1110.4043 (2011).


\bibitem{Myers1966}
W.D. Myers and W.J. Swiatecki, Nucl. Phys. \textbf{A81}, 1 (1966).

\bibitem{Moller1995} 
P. M\"oller, J.R. Nix and W.D. Myers and W.J. Swiatecki, Atomic Data and Nuclear Data Tables \textbf{59}, 185 (1995).

\bibitem{Pomorski2003}
K. Pomorski and J. Dudek, Phys. Rev. \textbf{C67}, 044316 (2003).


\bibitem{Akmal1998}
A. Akmal, V.R. Pandharipande and D.G. Ravenhall, Phys. Rev. \textbf{C58}, 1804 (1998).

\bibitem{Carlson2003}
J. Carlson, J. Morales, V.R. Pandharipande and D.G. Ravenhall, Phys. Rev. \textbf{C68}, 025802 (2003).

\bibitem{Schwenk2005}
A. Schwenk and C.J. Pethick, Phys. Rev. Lett. \textbf{95}, 160401 (2005).


\bibitem{Steiner2005} 
A.W. Steiner, M. Prakash, J.M. Lattimer and P.J. Ellis, Physics Reports \textbf{411}, 6, 325 (2005).

\bibitem{Ravenhall2004}
D.G. Ravenhall and C.J. Pethick, ApJ \textbf{424}, 2, 846 (2004).

\bibitem{Newton2009b}
W.G. Newton, Bao-An Li, Phys. Rev. \textbf{C80}, 065809 (2009).

\bibitem{Oyamatsu2007}
K. Oyamatsu and K. Iida, Phys. Rev. {\bf C75}, 015801 (2007).

\bibitem{Steiner2008} 
A.W. Steiner, Phys. Rev. \textbf{C77}, 035805 (2008).

\bibitem{Xu2009}
J. Xu, L.W. Chen, Bao-An Li and H.R. Ma, Phys. Rev. {\bf C79}, 035802 (2009); ApJ \textbf{697}, 1549 (2009).

\bibitem{FuturePaper} 
W.G. Newton, Bao-An Li, \emph{to be published}


\bibitem{Iida1997}
K. Iida and K. Sato, ApJ \textbf{477}, 294 (1997).

\bibitem{Watanabe2000} 
G. Watanabe, K. Iida and K. Sato, Nucl. Phys. \textbf{A676}, 455 (2000); G. Watanabe, K. Iida and K. Sato, Nucl. Phys. \textbf{A687}, 512 (2000).

\bibitem{LorenzPethick1993} 
C.P. Lorenz, D.G. Ravenhall and C.J. Pethick, Phys. Rev. Lett. \textbf{70}, 4, 379 (1993).


\bibitem{Ravenhall1983.2} 
D.G. Ravenhall, C.J. Pethick and J.M. Lattimer, Nucl. Phys. \textbf{A407}, 571 (1983).

\bibitem{Lattimer1985} 
J.M. Lattimer, C.J. Pethick, D.G. Ravenhall and D.Q. Lamb, Nucl. Phys. \textbf{A432}, 646 (1985).

\bibitem{Lorenz1991} 
C.P. Lorenz, PhD Thesis, University of Illinois.

\bibitem{Douchin2000} 
F. Douchin, P. Haensel and J. Meyer, Nucl. Phys. \textbf{A665}, 419 (2000).

\bibitem{Danielewicz2003}
P. Danielewicz, Nucl. Phys. \textbf{A727}, 233 (2003).



\bibitem{Ogata1990}
S. Ogata and S. Ichimaru, Phys. Rev. {\bf A42}, 8, 4867 (1990).

\bibitem{Strohmayer1991}
T. Strohmayer, H.M. van Horn, S. Ogata, H. Iyetomi and S. Ichimaru, ApJ \textbf{375}, 679 (1991).

\bibitem{Chugunov2010}
A.I. Chugunov and C.J. Horowitz, MNRAS \textbf{407}, 1, L54 (2010).

\bibitem{Chamel2008}
N. Chamel and P. Haensel, Liv. Rev. Rel., \textbf{11},  10 (2008).


\bibitem{PethickRev1995}
C.J. Pethick and D.G. Ravenhall, Annu. Rev. Nucl. Part. Sci. \textbf{45}, 429 (1995).


\bibitem{Nakazato2009}
K. Nakazato, K. Oyamatsu and S. Yamada, Phys. Rev. Lett. \textbf{103}, 132501 (2009).


\bibitem{RBP1972}
D.G. Ravenhall, C.D. Bennett and C.J. Pethick, Phys. Rev. Lett \textbf{28}, 15, 978 (1972).


\bibitem{Steiner2010}
A.W. Steiner, J.M. Lattimer and E.F. Brown, ApJ \textbf{722}, 1, 33 (2010).

\bibitem{Demo10} P. B. Demorest,  T. Pennucci, S. M. Ransom,  M. S. E. Roberts  and J. W. T. Hessels, Nature \textbf{467}, 1081 (2010).

\bibitem{Chamel2005b}
N. Chamel, Nucl. Phys. \textbf{A747}, 109 (2005).



\bibitem{mywebsite}
\url{http://williamnewton.wordpress.com/ns-eos}

\bibitem{Sandulescu2007}
C. Monrozeau, J. Margueron and N. Sandulescu, Phys. Rev. \textbf{C75}, 6, 065807 (2007).


\bibitem{Chamel2007}
N. Chamel, S. Naimi, E. Khan and J. Margueron, Phys. Rev. \textbf{C75}, 5, 055806 (2007).



\bibitem{Buchler1971}
J.-R. Buchler and Z. Barkat, ApJL \textbf{7} 167 (1971).

\bibitem{Oyamatsu1993}
K. Oyamatsu, Nucl. Phys. \textbf{A561}, 431 (1993).


\bibitem{Cheng1997}
K.S. Cheng, C.C. Yao and Z.G. Dai, Phys. Rev. \textbf{C55}, 4, 2092 (1997).


\bibitem{Onsi2008}
M. Onsi, A.K. Dutta, H. Chatri, S. Goriely, N. Chamel and J.M.Pearson, Phys. Rev. \textbf{C77}, 065805 (2008).


\bibitem{Negele1973}
J.W. Negele, and D. Vautherin, Nucl. Phys. \textbf{A207}, 298 (1973).

\bibitem{Montani2004}
F. Montani, C. May, and H. M\"uther, Phys. Rev. \textbf{C69}, 065801 (2004).

\bibitem{Baldo2005}
M. Baldo, U. Lombardo, E.E. Saperstein, and S.V. Tolokonnikov, Nucl. Phys. \textbf{A750}, 409 (2005).


\bibitem{Magierski2002}
P. Magierski and P.-H. Heenen, Phys. Rev. \textbf{C65}, 045804 (2002). 

\bibitem{Gogelein2008}
P. G\"ogelein, E.N.E. van Dalen, C. Fuchs and H. M\"uther, Phys. Rev. \textbf{C77}, 025802 (2008).

\bibitem{Newton2009}
W.G. Newton and J.R. Stone, Phys. Rev. \textbf{C79}, 055801 (2009).


\bibitem{Maruyama1998}
T. Maruyama, K. Niita, K. Oyamatsu, T. Maruyama, S. Chiba and A. Iwamoto, 1998, Phys. Rev. \textbf{C57}, 655 (1998).

\bibitem{Horowitz2004}
C.J. Horowitz, M.A. Perez-Garcia and J. Piekarewcz, J., Phys. Rev. \textbf{C69}, 045804 (2004).

\bibitem{Watanabe2001}
G. Watanabe, K. Iida and K. Sato, Prog. Th. Phys. \textbf{106}, 551 (2001).

\bibitem{Sonoda2007}
H. Sonoda, G. Watanabe, K. Sato, T. Takiwaki, K. Yasuoka and T. Ebisuzaki, Phys. Rev. \textbf{C75}, 042801 (2007).


\end{thebibliography}
\end{document}